\documentclass[a4paper,11pt]{article}
\pdfoutput=1 
\usepackage{jheppub} 
\usepackage{epsfig}
\usepackage{caption}
\usepackage{subcaption}
\usepackage{color,soul,bm}
\usepackage{float}
\usepackage{amsfonts}
\usepackage{url}
\usepackage{slashed}
\usepackage{accents}
\usepackage{lipsum}
\usepackage{mathtools}
\usepackage{physics}

\usepackage[normalem]{ulem}

\hypersetup{
  bookmarks=true,         
  unicode=false,          
  pdftoolbar=true,        
 pdfmenubar=true,        
 pdffitwindow=true,     
 pdfstartview={FitH},    
 pdfsubject={Dark Matter},   
 pdfnewwindow=true,      
 pdfcreator={RevTeX},
 colorlinks=true,       
 linkcolor=red,          
 citecolor=blue,        
 filecolor=black,      
 urlcolor=blue,           
  }
\title{From Non-Thermal to Thermal: Cosmological Imprints of Self-Interacting Secluded Dark Matter  }
\author[a]{Koushik Dutta,}
\author[b]{Deep Ghosh}
\author[a]{and Ahamadullah Khan}
\affiliation[a]{Department of Physical Sciences, Indian Institute of Science Education and Research (IISER) Kolkata, Campus Road, Mohanpur, West Bengal 741246.}
\affiliation[b]{Theory Division, Saha Institute of Nuclear Physics (SINP),
Sector-I, Block-AF, Bidhannagar, Kolkata 700064.}
\emailAdd{koushik@iiserkol.ac.in}
\emailAdd{matrideb1@gmail.com}
\emailAdd{ak22rs019@iiserkol.ac.in }

\abstract{Dark matter (DM), secluded from the Standard Model (SM), can attain internal thermalization through self-interactions. We investigate the thermalization of a real scalar DM field with a quartic self-coupling, initially produced from the perturbative decay of the coherently oscillating inflaton field during reheating. The thermalization process has been studied numerically by solving Boltzmann equation of the DM momentum distribution function in presence of $2\rightarrow2$ DM elastic self-scatterings. In the absence of number-changing processes, the thermalized dark sector is characterized by a temperature and a nonzero chemical potential, determined by the inflaton mass and the reheat temperature. We find that \textit{if the inflaton mass is greater (smaller) than the reheat temperature, the DM temperature can be greater (smaller) than the SM temperature during Big Bang Nucleosynthesis, without violating any cosmological constraints}. This is realized with a negative (positive) chemical potential for the DM temperature greater (smaller) than the SM one. Depending on DM production history, we can achieve the lower limit on the DM mass much below than in \textit{warm DM} scenarios. The thermalization via self-scatterings tends to suppress the DM co-moving free-streaming length, compared to the non-thermal scenario without any collisions. Consequently, the lower limit on the DM mass weakens approximately by $40\%$ for a suitable choice of quartic self-coupling than in the non-thermal scenario. We obtain the lower limits on the thermal DM mass to be $2.2\times10^{-10}$ GeV and $0.47$ GeV for reheat temperatures, $5\times 10^6$ GeV and $5$ MeV respectively, while the inflaton mass is set at $10^3$ GeV.}

\begin{document} 
\maketitle
\flushbottom  
\section{Introduction and Summary}
Despite overwhelming evidences of dark matter (DM) through gravitational interaction, signature of its non-gravitational interactions remain elusive till date. The large scale structure formation can be completely determined given the DM density and the \textit{free-streaming length} (Jeans length for collisional DM) at the time of the matter-radiation equality epoch \cite{Rubakov:2017xzr}. However, the generation of required  DM density and the free-streaming length from some initial conditions in the early universe are not yet settled, thereby every possible idea is needed to be explored, particularly involving non-gravitational interactions of particle DM. This is also important in terms of devising experimental methodologies and observational strategies to detect DM micro-physics beyond its gravitational footprints.

In this regard, DM scenarios can broadly be classified into two categories, namely \textit{thermal} and \textit{non-thermal}, depending on the shape of its initial phase-space distribution in the early universe. For thermal DM, the initial phase-space distribution takes the equilibrium form according to the spin of DM particles, for which the initial energy density is completely determined by two thermodynamic parameters, namely, temperature and chemical potential.
The classic example in this context is the WIMP DM scenario \cite{Lee:1977ua,Steigman:1984ac}, in which DM is in thermal equilibrium with the Standard Model (SM) plasma in the early universe. The DM temperature is identical with the SM one, while the chemical potential is taken as zero. As a consequence, the relic density of DM is miraculously achieved with DM annihilation to SM states, whose rate is fixed by the weak scale coupling, rendering DM mass within GeV to TeV range. In fact, the upper limit on the DM comes out  to be around 100 TeV from S-matrix unitarity of DM annihilation rate \cite{Griest:1989wd,Hui:2001wy,Bhatia:2020itt}, whereas the lower limit is around a MeV from the observations of Big Bang Nucleosynthesis (BBN). To emphasize, these constraints on the DM parameters are valid only for vanishing chemical potential, which is itself an assumption rather than a general result. If we allow non-zero chemical potential of DM, signifying a particle-antiparticle asymmetry in the dark sector, the DM relic density can be determined in terms of primordial asymmetry, rather than the DM annihilation rate as in the WIMP scenario \cite{Petraki:2013wwa,Zurek:2013wia,Ghosh:2020lma}. As a result, the unitarity limit on the DM mass is modified to 15 GeV as shown in Ref.\cite{Ghosh:2020lma}. 

In the above two scenarios, DM particles reach  thermal equilibrium with the SM thermal bath through DM-SM interactions. However, non-observation of such DM scenarios in the direct, indirect and collider experiments, has put the paradigm under some challenges, despite recent hints from LUX-ZEPLIN experiments \cite{LZ:2026axp,Pospelov:2026ewn,Bose:2026ndd}.  As an alternative, internally thermalized DM scenarios have been proposed, in which DM is thermalized via self-scattering in the absence of sufficient DM-SM interaction \cite{Pospelov:2007mp,Feng:2008mu,Hambye:2019dwd}. Although the DM self-scattering is yet to be confirmed, the merging of Bullet clusters put the upper bound on the DM self-scattering cross-section as, $\sigma_{\rm DM}\sim 1 ~{\rm cm^2} \left(m_{\rm DM}/gm\right)$ \cite{Harvey:2015hha,Tulin:2017ara}, which leaves ample room for a sizable self-scattering rate to facilitate DM thermalization. In these scenarios, the DM thermodynamics is characterized by a temperature, independent of the SM one and also by a chemical potential, determined by DM scatterings within itself. Evidently, the DM relic density is determined by the inelastic scattering within the DM sector. The simplest possibility is $3~{\rm DM}\rightarrow  2 ~{\rm DM}$ scattering, for which the DM chemical potential becomes zero. In this scenario, the unitarity bound on the DM mass becomes strongly dependent on the initial DM temperature, which is a free parameter of the theory. For example, the upper bound on the DM mass can vary within $80 ~\rm eV$ to $700$ TeV for an allowed range of initial DM temperatures, $T_{\rm SM}/9100 \lesssim T_{\rm DM}\lesssim T_{\rm SM}/1.1$ \cite{Ghosh:2022asg}. These results are expected to be modified in presence of a non-vanishing chemical potential. 

It is evident that the history of the non-thermal phase becomes irrelevant for a thermal DM candidate, given the initial DM temperature and the chemical potential in general. For WIMP-like scenarios, the situation is further simplified since the DM temperature is identical with the SM one, whereas for the so-called \textit{secluded} DM scenarios, the DM temperature and chemical potential are free parameters, potentially sensitive to the non-thermal phase. Therefore, for a secluded DM scenario, it becomes important to know the entire history of DM thermalization, which is going to be the main focus of the current study.

As an aside, DM is non-thermal if its phase-space distribution do not take any equilibrium form for the entire cosmic history, due to insufficient interaction with SM particles or with itself. In this situation, DM temperature and chemical potential are not well defined. As a result, the exact evolution of its phase-space distribution function is required to realize the relic density and free-streaming length of DM particles \cite{Ghosh:2022hen}. 

In this article, we consider a real scalar DM with quartic self-interaction term, whose initial density is set by the perturbative decay of a inflaton condensate at the time of reheating. At the beginning, the phase space density of DM particles is non-thermal as the momentum distribution function is peaked at the half of the inflaton mass, for a two-body decay. We study the evolution of DM momentum distribution function in presence of DM elastic self-scatterings facilitated by the quartic self-interaction to assess the thermalization of nascent DM particles, followed by determination of the DM temperature and the chemical potential. In particular, we determine the time of thermalization as a function of DM self-interaction for a given inflaton mass ($m_\phi$) and the reheat temperature ($T_R$). In the absence of any other number-changing process in the dark sector, the DM number is fixed at the very outset from the inflaton decay, thereby the DM number density only redshifts due to Hubble expansion. As a consequence, DM acquires a non-zero chemical potential to conserve the total DM number, unlike the cannibal DM scenarios in which inelastic scatterings renders chemical potential to vanish. Therefore, the constraints on the DM mass in our scenario are governed by both the DM temperature and chemical potential. The main findings of our study are summarized below:
\begin{enumerate}
    \item The initial DM temperature for sufficiently large quartic self-coupling is set by the inflaton mass ($m_\phi$), i.e. $T^i_{\rm DM}=m_\phi/6$. In the relativistic regime the dark sector, can be hotter (colder) than the SM one for $T_R < m_\phi/6$ ($T_R> m_\phi/6$). In the absence of number-changing processes, the ratio of DM temperature to the SM one is completely determined by $m_\phi$ and $T_R$.
    \item The DM chemical potential becomes correlated with the DM temperature and the DM mass since the total DM number is fixed at the production. The DM chemical potential can take both positive and negative values depending on reheat temperature and inflation mass. We show for a high reheat scenario with $T_R=5\times 10^6$ GeV, the DM chemical potential is always positive, whereas for a low reheat scenario with $T_R=5$ MeV, it takes negative value in the relativistic regime. In both cases, we set $m_\phi = 10^3$ GeV.
    
    \item A hotter dark sector realized in the low reheat scenario is found to be completely allowed from the constraint on $N_{\rm eff}$ at the time of BBN. A negative chemical potential suppresses the DM energy density so that all cosmological constraints are satisfied. This is to be contrasted with most of the secluded sector DM scenarios \cite{Pospelov:2007mp,Feng:2008mu,Hambye:2019dwd}, which often preach a DM temperature smaller than the SM one. 
    \item The lower limits on the DM mass from the BBN and Cosmic Microwave Background (CMB) radiation turn out to be independent of self-interaction strength. This stems from the fact the DM temperature is set by the inflaton mass and the reheat temperature in presence of elastic scatterings, thereby hiding the thermalization history.  
    \item We find that lower limit of the DM mass is well below the warm DM scenarios for high reheat scenario from CMB and BBN observations. In particular, the lower limits are found to be $\sim 10^{-12}$ GeV and $\sim 10^{-3}$ GeV for high and low reheat scenarios respectively. 
    \item The effect of elastic scattering, in turn the history of thermalization becomes apparent in the DM co-moving free-streaming length. The lower bound on the DM mass originating from the free-streaming length weakens approximately by $40\%$ due to elastic scatterings. The bound is explicitly dependent on the self-interaction strength, unlike the previous constraints from BBN and CMB observations.
     
\end{enumerate}
This paper is organized as follows. In Sec.~\ref{sec:sec2} we demonstrate DM production from the inflaton decay and subsequent thermalization via self-scatterings. We also discuss the evolution of DM temperature and chemical potential. In Sec.~\ref{sec:sec3}, we discuss cosmological constraints from BBN, CMB and free-streaming length for our secluded DM scenario, followed by concluding remarks in Sec.~\ref{sec:sec4}. We have also two appendices to explain the numerical techniques of solving Boltzmann equation and details of thermalization process.  

\section{Scalar dark matter production and its thermalization}
\label{sec:sec2}
In this section, we discuss the perturbative production of DM through inflaton decays and its subsequent thermalization via elastic scattering among DM particles. We assume that the DM sector has no direct interactions with the SM sector, thereby constituting a \textit{secluded} dark sector. Thermalization establishes a thermal distribution characterized by two thermodynamic quantities: the temperature and the chemical potential. The DM temperature can be either higher or lower than the SM temperature, with a non-zero chemical potential, which are sensitive to pre-thermal history of DM production.

\subsection{DM production from the inflaton decay}
During the reheating, the inflaton field coherently oscillates and dissipates its energy mostly to the visible sector, initiating the radiation-dominated phase of the universe. During this period, a small fraction of energy can populate the dark sector as well, which is well-studied in  the literature \cite{Moroi:1994rs, Kawasaki:1995cy, Moroi:1999zb, Chung:1998zb, Chung:1998ua, Jeong:2011sg, Ellis:2015jpg, Harigaya:2014waa, Drees:2021lbm, Garcia:2018wtq, Harigaya:2019tzu, Garcia:2020eof,  Garcia:2020wiy, Moroi:2020has,Adshead:2016xxj,Hardy:2017wkr,Berlin:2016gtr,Adshead:2019uwj,March-Russell:2020nun, Bhattacharya:2020zap, Haque:2022kez, Haque:2021mab, Ghosh:2022hen,Garcia:2022vwm, Ghoshal:2022jeo, Kaneta:2022gug, Ghoshal:2022ruy,Ghosh:2023mnh}. The dissipation of inflaton energy can happen via a brief non-perturbative phase, known as preheating~\cite{Garcia:2022vwm,Kofman:1997yn}. However, here we only focus on the perturbative decay of the inflaton field, a spin-$0$ particle in its zero momentum state, producing a pair of daughter particles with momenta $p= m_{\phi}/2$ at the reheating. In addition, we consider the inflaton potential during reheating to be quadratic, which makes the inflaton decay rates (and the inflaton mass) to be time-independent \cite{Garcia:2020eof}. Under these assumption, we consider a real scalar DM field, $\chi$, charged under a $\mathcal{Z}_2$ symmetry. The DM production and its subsequent thermalization can be realized with a trilinear interaction with the inflaton field, $\phi$ and a quartic self-interaction term, given by  
\begin{align}
-\mathcal{L}_{\rm int} \supset  \frac{\beta}{2} \phi \chi^2 + \frac{\lambda_\chi}{4 !} ~\chi^4.
\label{eq:theory}
\end{align}

The DM production happens through the inflaton decay to two DM particles. The partial decay width is given by
\begin{align}
\Gamma_{\phi \rightarrow \chi \chi} = \frac{\beta^2}{32\pi m_\phi} \sqrt{1-\frac{4m^2_\chi}{m^2_\phi}},
\end{align}
where $m_\phi$ and $m_\chi$ are the inflaton mass and the DM mass respectively. In the instantaneous reheating approximation, the DM number density at the reheat temperature ($T_R$) becomes
\begin{align}
n_\chi (T_R)= 2 \frac{\rho_\phi(T_R)}{m_\phi} {\rm Br} (\phi \rightarrow \chi \chi),
\label{eq:den}
\end{align}
where the branching fraction, ${\rm Br} (\phi \rightarrow \chi \chi) \approx \Gamma_{\phi \rightarrow \chi \chi}/\Gamma_{\phi \rightarrow {\rm SM SM}} $. The decay width to the visible sector, $\Gamma_{\phi \rightarrow \rm SM SM}$ dominates the total decay width of the inflaton. In the instantaneous reheating approximation, $\Gamma_{\phi \rightarrow {\rm SM SM}}$ is traded with the Hubble parameter at reheating, i.e. $H(T_R)= \Gamma_{\phi \rightarrow {\rm SM SM}} $. This also implies that the inflaton density, $\rho_\phi$ is approximately equal to the radiation density ($\rho_R$) at reheating, i.e. $\rho_\phi(T_R) \approx \rho_R (T_R)= \frac{\pi^2}{30}g_*(T_R) T^4_R $. Using all these inputs, the DM number density is found to be 
\begin{align}
n_\chi (T_R)= \frac{1}{16\sqrt{10}}\left(\frac{\beta}{m_\phi}\right)^2 \sqrt{M^2_P ~ g_*(T_R)\left(1-\frac{4m^2_\chi}{m^2_\phi}\right)} ~~ T^2_R ,
\label{eq:num}
\end{align}
where $M_P ~(=2.43\times 10^{18}~{\rm GeV})$  is the reduced Planck mass and $g_*(T_R)$ is the effective degrees of freedom at reheating. To note, $T_R$ can be calculated  given the interaction between the inflaton and the SM fields. In Ref.~\cite{Ghosh:2023mnh}, $T_R$ is calculated from similar cubic coupling of the inflaton to the SM Higgs particle,  as follows :

\begin{align}
T_R \simeq \left(\frac{90}{\pi^2 g_{*}}\right)^{1/4} \sqrt{\Gamma_{\phi \rightarrow H^\dagger H} M_P}.
\end{align}

For our subsequent discussion, we will consider $T_R$ as a free parameter without referring to the inflaton-Higgs coupling. In addition, we assume instantaneous thermalization of the visible sector as the SM interactions are sufficient enough for setting a thermal bath within a short timescale, compared to that of the dark sector. Furthermore, there is no additional DM number-changing process, other than the inflaton decay, as apparent from Eq.~\ref{eq:theory}. Consequently, the co-moving number density $Y_\chi$ ($= n_\chi /s, s$ being the entropy density of the SM bath) remains constant since the DM production. Hence, the DM relic density is also set by initial conditions at the time of reheating as the following  
\begin{align}
 \Omega_\chi h^2 \simeq 0.12
\left(\frac{5\times 10^6~ {\rm GeV}}{T_R}\right)\left(\frac{10^3~{\rm GeV}}{m_\phi}\right)^2\left(\frac{m_\chi}{10^{-3}~{\rm GeV}}\right) \left(\frac{\beta}{1.4\times 10^{-5} ~{\rm GeV}}  \right)^2 \left(\frac{106.25}{g_{*}(T_R)}  \right)^{1/2},
\label{eq:relic}
\end{align}
where, $\Omega_\chi = m_\chi s_0 ~Y_\chi (T_R)/\rho_c$. $s_0$ is the entropy density and $\rho_c$ is the critical density of the current universe. In our setup, we have five independent parameters to start with, namely, $T_R,m_\phi, m_\chi, \beta$ and $\lambda_\chi$. Apart from $\lambda_\chi$, all other parameters are related to DM production. Given the DM relic density in Eq.\ref{eq:relic}, we eliminate $\beta$. Throughout this paper, we fix $m_\phi$ to be $10^3\, \rm GeV$ and choose two benchmark reheat temperatures: $T_R = 5\times10^6 \, {\rm GeV}$ and $T_R = 5\, {\rm MeV}$. Hence, we left with $m_\chi$ and $\lambda_\chi$ as free parameters, which will be constrained from cosmological and astrophysical observations.

In short, we consider the inflaton to decay predominantly to the SM sector which thermalizes instantaneously producing initial temperature, $T_R$, and at the same time some fraction of energy density goes to the dark sector at the typical timescale of $\Gamma^{-1}_{\phi\rightarrow \chi\chi}$. This dark sector has no direct interaction with the SM sector. Our broad goal is to study thermalization of DM particles and its cosmological implications in presence of self-interaction.

\subsection{DM Thermalization via self-interaction}
 In this subsection, we will discuss the evolution of DM momentum distribution in presence of DM elastic scattering from an initial non-thermal phase set at its production. In particular, we solve the Boltzmann equation of the DM distribution function for specific values of different parameters to show how thermalization proceeds.
 
 The initial DM momentum distribution is determined by the instantaneous decay of coherently oscillating inflaton field at the reheating and takes the following form :
\begin{align}
f_{\chi}^{\rm ini} (p,t_{\rm ini}=t(T_R)) = A ~\delta (p-m_\phi/2)~~~ \text{for}~~ m_\chi \ll m_\phi,
\label{eq:ini}
\end{align} 
where $p$ corresponds to the physical momentum and the proportionality constant, $A$, is calculated from the initial DM number density shown in Eq.\ref{eq:num} :  

 \begin{align}
A &= (2\pi)^3 \frac{n_\chi(T_R)}{\int d^3 p\, \delta(p-m_\phi/2)}\nonumber \\
&= 1.86\times 10^{11}
\left(\frac{10^3\,{\rm GeV}}{m_\phi}\right)^4
\left(\frac{T_R}{5\times10^6\,{\rm GeV}}\right)^2
\left(\frac{\beta}{1.4\times10^{-5}\,{\rm GeV}}\right)^2
\left(\frac{g_*(T_R)}{100}\right)^{1/2}.
\label{eq:norm}
\end{align}
For \textit{sufficient} elastic scattering, the DM momentum distribution eventually takes the equilibrium form, i.e. Maxwell-Boltzmann form for our case, ignoring the quantum effects. Once, the dark sector is thermalized, its thermodynamics is completely determined by its temperature and chemical potential in general. The transition from the non-thermal phase to thermal phase is determined by the interplay of the Hubble expansion rate and $2\rightarrow 2$ DM self-scattering rate governed by $\frac{\lambda_\chi}{4!}\chi^4$ term as well as $2\rightarrow 2$ DM scattering rate mediated by inflaton field governed by $\beta \phi \chi^2$ term. In fact, the latter term brings correction to $\lambda_\chi$, proportional to $(\beta/m_\phi)^2$. For simplicity, we take $\lambda_\chi$ as the effective quartic self-interaction strength without explicitly mentioning the correction further in the text.

The thermodynamic temperature and the chemical potential are defined only after the thermalization, therefore the determination of thermalization timescale is important. To determine that, the evolution of DM momentum distribution function in presence of DM elastic collisions, is tracked using the following Boltzmann equation:
\begin{align}
\frac{\partial f_\chi(p_1,t)}{\partial t}- H p_1\frac{\partial f_\chi (p_1,t)}{\partial p_1}= \, \mathcal{C}[f_\chi(p_1, t)]  & = -\frac{1}{2 E_1} \int \prod^4_{i=2} \frac{d^3 p_i}{(2 \pi)^3 2 E_i}(2 \pi)^4 \delta^4\left(p_1+p_2-p_3-p_4\right)\nonumber\\
&|M|^2 \bigg[f_\chi (p_1) f_\chi (p_2) - f_\chi(p_3) f_\chi(p_4)\bigg],
\label{eq:boltz}
\end{align}
where, the probability amplitude of the DM self-scattering, $|M|^2 = \lambda^2_\chi$ and the energy of the DM particle, $E_i=\sqrt{p^2_i+m^2_\chi}$. Now, we need to solve for $f_\chi (p_1,t)$ with the initial condition shown in Eq.\ref{eq:ini} and a choice of self-interaction strength, $\lambda_\chi$ and $m_\chi$. The main challenging task in solving this integro-differential equation is to calculate the collision term ($C[f_\chi]$) in the right hand side of the equation, which can be reduced to a two-dimensional integration from the nine-dimensional integration for a constant $|M|^2$ \cite{Du:2021jcj,Ala-Mattinen:2019mpa,Bringmann:2006mu}. For details see Appendix \ref{app:boltz}. Under the instantaneous reheating approximation, the universe becomes radiation-dominated since the DM production, thereby we track the evolution of the DM momentum distribution function in the radiation-dominated universe. 
\begin{figure}[t]
\centering
\includegraphics[scale=0.4]{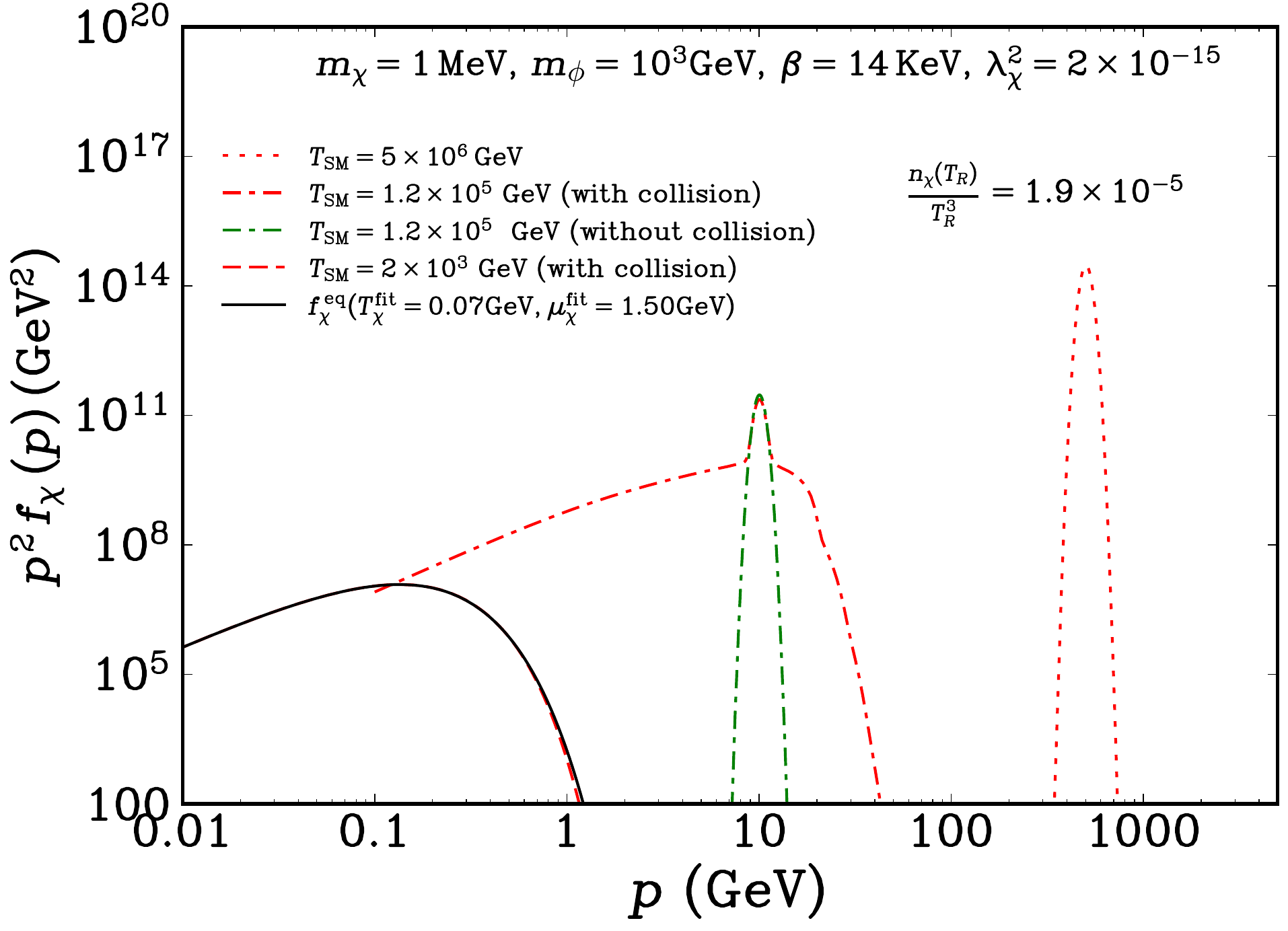}
\caption{Momentum distribution function of DM at different SM bath temperatures, showing the effect of elastic scatterings between DM particles which produced from the inflaton decay at the reheat temperature, $T_R=5\times 10^6 $ GeV. We show how an initial non-thermal distribution (\textit{red dotted line}) is evolved nearly into a Maxwell-Boltzmann equilibrium distribution (\textit{red dashed line}) due to momentum redistribution via quartic self-interaction of DM. For clarity, we also show the distribution function in an intermediate epoch with (\textit{red dot-dashed line}) and without collision (\textit{green dot-dashed line}). }
\label{fig:dist}
\end{figure}
In Fig.\ref{fig:dist}, we show the DM momentum distribution functions in three different epochs, obtained numerically after solving Eq.\ref{eq:boltz} in Python using ``BDF" ODE solver. For illustration, we choose benchmark points as $m_\chi= 1~{\rm MeV}$, $T_R=5\times 10^6$ GeV and $m_\phi = 10^3~ {\rm GeV}$, which fixes $\beta$ to be $14$ keV from the relic density constraint, Eq.~\ref{eq:relic}. As is evident, in the absence of any other number-changing process, DM particles only redistribute their momenta from its initial form (shown in red dotted line) due to elastic scatterings among themselves, whose rate is determined by both $\lambda_\chi$ and the number density. As a sanity check of our numerical code, we have calculated the co-moving number density of DM to be remained constant at its initial value, i.e. $n_\chi(T_R)/T^3_R = 1.9\times 10^{-5}$, throughout the evolution history.

The effect of elastic scattering is apparent from the DM momentum distribution function shown by the red dot-dashed line at SM radiation bath temperature, $T_{\rm SM}=1.2\times 10^5$ GeV. It should be compared with the initial distribution function and its red-shifted version shown by the green dot-dashed line. Due to cosmic expansion, all physical momenta red-shifts as, $p\propto 1/a(t)$, $a(t)$ being the usual scale factor. Consequently, the initial momentum distribution maintains its shape in the absence of any collision, while the number density dilutes as, $n_\chi(t) \propto 1/a(t)^3$. In presence of elastic collisions, each momenta is modified for a combined effect of collisions and the cosmic expansion. As a result, the DM distribution deviates from its initial shape as shown by red dot-dashed line in Fig.\ref{fig:dist}. To note, elastic scatterings populate lower momentum states, thereby tending towards the Maxwell-Boltzmann (MB) distribution, i.e. $f_\chi \propto e^{-(E-\mu_\chi)/T_\chi}$. Eventually, at $T_{\rm SM}= 2\times 10^3$ GeV, the DM distribution (shown in red dashed line) matches well with the MB distribution with temperature $T^{\rm fit}_\chi = 0.07$ GeV and chemical potential, $\mu^{\rm fit}_\chi=1.50$ GeV, depicted by the black solid line in Fig.\ref{fig:dist}. To note, DM particles do not undergo any number-changing process after instantaneous production from the inflaton decay, thereby requiring non-zero chemical potential, unlike the standard WIMP scenario with a vanishing chemical potential. In this case, a non-zero chemical potential does not signify any particle-antiparticle asymmetry \cite{Lin:2023fao}, rather it signifies the conservation of DM number in elastic scatterings within the dark sector. 

\subsection{Evolution of DM temperature and chemical potential }
As a follow-up of the previous subsection, we will now discuss the cosmological evolution of DM temperature and chemical potential. We will derive the results analytically first and will also cross-check the same by solving Boltzmann equation numerically. In the end, we will argue that the secluded sector dark matter can be at both higher and lower temperature compared to the Standard Model sector.

Given a distribution function $f_\chi (p,t)$, we define the following quantity called the \textit{effective} DM temperature as \cite{Ghosh:2022asg,Bringmann:2006mu} 
\begin{align}
T^{\rm eff}_\chi(t) = \frac{1}{n_\chi} \int \frac{d^3 p}{(2\pi)^3} \frac{p^2}{3E} f_\chi (p,t).
 \label{eq:temp}
 \end{align}
For an equilibrium distribution function i.e. $f^{\rm eq}_{\chi} (p, T_\chi, \mu_\chi)$, the \textit{effective} DM temperature $T^{\rm eff}_\chi(t)$ is evaluated to be the thermodynamic temperature, $T_\chi(t)$. However, for non-thermal case, $T^{\rm eff}_\chi(t)$ can only be interpreted as average kinetic energy without the notion of thermodynamic temperature $T_\chi(t)$. To be precise, $T^{\rm eff}_\chi(t) = \expval{E}/3$ in the relativistic limit, where $\expval{E}$ being the average energy per particle and $T^{\rm eff}_\chi(t) = \frac{2}{3}\expval{p^2/2m_\chi}$ in the non-relativistic limit. The same two expressions hold for the thermal case as well.

In our case, the DM number density and energy density are separately conserved up to the trivial redshift from Hubble expansion. Therefore, in the relativistic regime, $\expval{E}$ will only redshift from its initial value, $\expval{E} = \expval{E}_{\rm ini}/a$. This immediately gives $T^{\rm eff}_\chi(t) = T^{\rm eff}_{\chi, {\rm ini}}/a$ in the same regime, where $T^{\rm eff}_{\chi, {\rm ini}} = \expval{E}_{\rm ini}/3$. Therefore, for a given a initial condition, i.e. $\expval{E}_{\rm ini}$, we can calculate $T_\chi^{\rm eff} (t)$ at any point of time, irrespective of the form of the distribution function during its evolution. In particular, the value of $T_\chi^{\rm eff} (t)$ is same for both the thermal and non-thermal distribution function except for the interpretation that for the thermal case it is thermodynamic temperature, whereas for the non-thermal one it is only the measure of average kinetic energy as argued in the previous paragraph.  

In our scenario, the initial condition comes from the inflaton decay, for which we start with a non-thermal distribution, characterised by the inflaton mass, $m_\phi$ and the reheat temperature, $T_R$, given in Eq.\ref{eq:ini}. Hence, $T^{\rm eff}_\chi(T_{\rm SM})$ in the \textit{relativistic} regime, is given by,
\begin{align}
T^{\rm eff}_\chi(T_{\rm SM}) =
\frac{\rho_\chi (T_R)}{3 n_\chi(T_R)} \frac{a(T_R)}{a(T_{\rm SM})}= \frac{m_\phi}{6T_R}\left(\frac{g_{*S}(T_{\rm SM})}{g_{*S}(T_R)}\right)^{1/3} T_{\rm SM},
\label{eq:tchirel}
\end{align} 
where $g_{*S}(T_{\rm SM})$ is the effective degrees of freedom contributing to the entropy density of the SM radiation bath. Note that, we use $T_{\rm SM}$ as a proxy for time. In deriving the above formula, we have used the conservation of total entropy density of the SM radiation, i.e. $s(T)a(T)^3=\frac{2\pi^2}{45}g_{*S}(T)T^3a(T)^3=\rm const$. Furthermore, the ratio of the DM energy density and number density is evaluated using Eq.\ref{eq:den}. It is evident from the above expression that the effective DM temperature ( also the thermodynamic temperature, $T_\chi(t)$, if it thermalizes) can be larger than that of the SM radiation bath if $\frac{m_\phi}{6T_R}\left(\frac{g_{*S}(T_{\rm SM})}{g_{*S}(T_R)}\right)^{1/3} > 1$.

In the \textit{non-relativistic} regime, the DM temperature varies as $1/a^2$, apparent from the redshift of $\expval{p^2/2m}$. We assume an instantaneous transition from relativistic to non-relativistic phase to find the scale factor, $a_{\rm NR}$, at which $T^{\rm eff}_\chi$ $= m_\chi/3$, taken as the marker of the onset of the non-relativistic regime. Therefore, the effective DM temperature can again be calculated in terms of the initial condition as follows :
\begin{align}
    T^{\rm eff}_{\chi}(T_{\rm SM}) = \frac{m^2_\phi}{12 m_\chi}\left(\frac{T_{\rm SM}}{T_R}\right)^2 \left(\frac{g_{*S}(T_{\rm SM})}{g_{*S}(T_R)}\right)^{2/3},
    \label{eq:tchinon}
\end{align}
where we have used, $a_{\rm NR}=(m_\phi/2m_\chi) a(T_R)$. 

Once the DM thermalizes, the normalization factor, $A$ in Eq.\ref{eq:norm} for the initial non-thermal distribution translates into  \textit{chemical potential} in the thermal distribution such that the correct DM number density is reproduced. This stems from the fact that the DM temperature decreases as the Universe expands. To maintain the equilibrium form with correct number density, we need to introduce the DM chemical potential in presence of number-conserving processes. The chemical potential in the relativistic regime can be written as,
 \begin{align}
     \mu_\chi (T_{\rm SM}) = T_\chi (T_{\rm SM})\log\left(\frac{\pi^2 n_\chi (T_{\rm SM})}{T^3_\chi(T_{\rm SM})}\right) = T_\chi(T_{\rm SM}) \log \left(\frac{1.9\times 10^{-9}~\rm GeV}{m_\chi} ~\frac{T^3_{\rm SM}}{T^3_\chi(T_{\rm SM})}g_{*S}(T_{\rm SM})\right),
     \label{eq:chem}
 \end{align}
where $n_{\chi} (T_{\rm SM})$ is calculated from the initial number density that satisfies the observed DM relic density, i.e. $n_\chi (T_R)= \frac{\Omega_\chi \rho_c}{s_0 m_\chi} s(T_R)$. For the expression of $\mu_\chi$, we have used the Maxwell-Boltzmann form of the DM momentum distribution. It is worth noting that the DM chemical potential can be either positive or negative, depending on the argument of the logarithmic function in Eq.~\ref{eq:chem}. Specifically, $\mu_\chi$ is greater (lesser) than zero if $m_\chi$ is lesser (greater) than a critical value, i.e. $ m_{\rm crit}=1.9\times10^{-9} {~\rm GeV} g_{*S}(T_{R})\left(6T_R/m_\phi\right)^3$. For a detailed discussion on the implication of a negative chemical potential see Sec.\ref{sec:sec3}. 

In the non-relativistic regime, i.e. $m_\chi \gg T_\chi$, the equilibrium  number density, $n_{eq} (t)\propto (m_\chi T_{\chi})^{3/2}~e^{\frac{\mu_\chi-m_\chi}{T_\chi}}$ and $T_{\chi} \propto 1/a^2(t)$, which prompts $\mu_\chi$ to approach the DM mass. Hence, in the non-relativistic regime, the DM chemical potential becomes a positive constant, irrespective of the temperature ratio between DM and SM sector. 

\begin{figure}[t]
\hspace{-0.8 cm}
\includegraphics[scale=0.64]{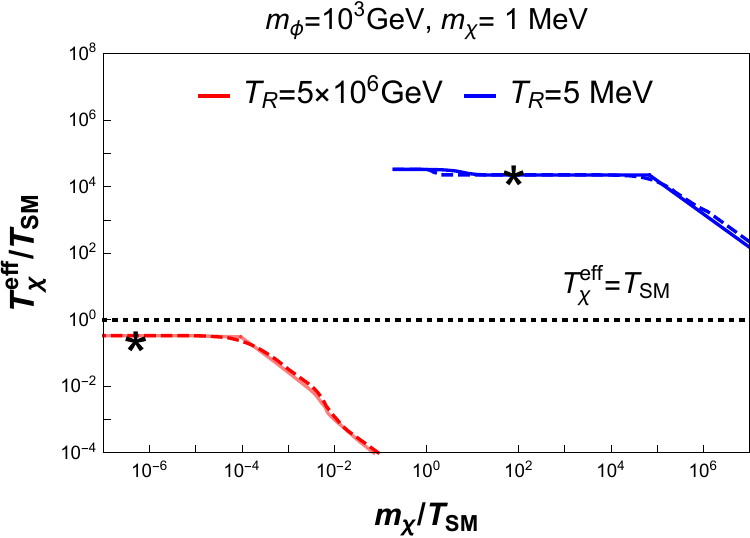}
\includegraphics[scale=0.63]{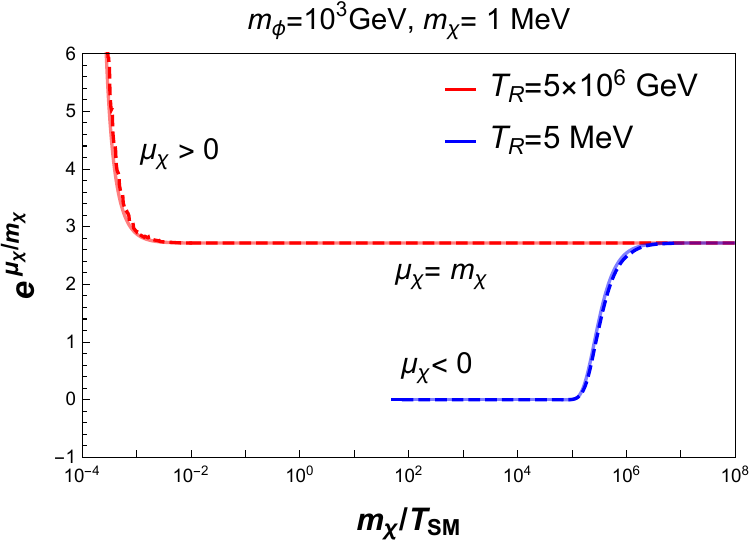}
\caption{\textit{Left panel} : Evolution of the temperature ratio ($T^{\rm eff}_\chi/T_{\rm SM}$) between the dark and visible sector for two benchmark reheat temperatures, $T_R=5\times 10^6$ GeV (red solid line) and $T_R=5$ MeV (blue line). For higher $T_R$, the DM sector is cooler than the visible one, whereas for lower $T_R$ the dark sector is hotter for a significant time in the cosmic history. We have scaled the temperature ratio of higher $T_R$ by a factor $10^4$. We show the epoch at which $T^{\rm eff}_\chi =T_\chi$, by ($\star$) symbol.  \textit{Right panel} : Evolution of corresponding DM chemical potential ($\mu_\chi$). For both high and low $T_R$, in the non-relativistic the chemical potential equals the DM mass. For relativistic regime, the chemical potential is positive (negative) for high (low) $T_R$. Since, for low $T_R$, $T_\chi/T_{\rm SM} \sim 10^4$ to start with, the chemical potential takes high negative value to maintain the DM number at its initial value. For both figures, the dashed lines indicate same thermodynamic quantities, inferred from the numerically calculated DM momentum distribution.}
\label{fig:ratio}
\end{figure}
In the left panel of Fig.\ref{fig:ratio} we show the evolution of the temperature ratio of the dark and visible sector for $T_R=5\times 10^6$ GeV and $T_R=5$ MeV, while fixing DM mass, $m_\chi=1$ MeV. We compare the analytical estimates with exact numerical results derived from the evolution of DM momentum distribution function. For two cases, the production of DM particles happen at different epochs, therefore, the evolution of DM temperature and the chemical potential (in the right panel) start from different $m_\chi/T_{\rm SM}$ values. For the high reheat scenario, we have scaled the ratio by a factor of $10^4$ to show it with the low reheat scenario in the same figure. For numerical results, we have taken the self-scattering amplitudes as $\lambda_\chi^2 = 2\times10^{-15}$ for high $T_R$, $\lambda_\chi^2=2\times10^{-5}$ for low $T_R$, for which DM particles thermalize. Note that,  $T^{\rm eff}_\chi(t)$ becomes the thermodynamic temperature $T_\chi(t)$ at $m_\chi/T_{\rm SM} = 5\times 10^{-7}$ and $m_\chi/T_{\rm SM} = 79.94$ for high and low reheating scenarios respectively. We denote the epochs by a marker ($\star$) in the left panel of Fig.\ref{fig:ratio}. The thermodynamic temperature is also verified using the numerical fitting of Maxwell-Boltzmann distribution function as done in Fig.\ref{fig:dist}. The analytical estimates (solid lines in Fig.\ref{fig:ratio}) found in Eq.\ref{eq:tchirel}, Eq.\ref{eq:tchinon} and Eq.\ref{eq:chem} are in excellent agreement with the numerical results (dashed lines), with an error of less than $10\%$.  This shows that in presence of elastic collision, initial conditions fix average kinetic energy, irrespective of the form of the DM momentum distribution function. This is due to the fact both the co-moving number density and energy density remains constant in presence of elastic scatterings. If there is any number-changing process present within the dark sector or with the SM thermal bath, the dependence on initial conditions will be wiped out, thereby the form of the distribution will be important to determine the thermodynamic quantities.

In the relativistic regime, the temperature ratio remains constant, set by initial conditions, with a mild variation due to change in the effective degrees of freedom. The initial value of the temperature ratio is smaller (greater) than unity for higher (lower) reheat temperature, while $m_\phi$ is fixed at $10^3$ GeV.
Hence, a hotter dark sector can be realized in our case with inflaton mass greater than the reheat temperature, irrespective of $m_\chi$. However, $m_\chi$ has an important role in determining $\mu_\chi$, as stated earlier. This is apparent from the right panel of Fig.\ref{fig:ratio}, where we show the evolution chemical potential for two benchmark reheat temperatures. In the high-reheating scenario, $\mu_\chi$ is always positive since $m_{\rm crit}\sim 5\times10^{6}~\mathrm{GeV}$,  is kinematically inaccessible for the current choice of inflaton mass. In the low-reheating scenario, $\mu_\chi<0$, as $m_{\rm crit} \simeq 5\times 10^{-22}$ GeV. Since, the total DM number is conserved throughout cosmic history, the evolution of chemical potential is correlated with the temperature evolution. Therefore, the chemical potential saturates to the DM mass irrespective of reheat scenarios in in the non-relativistic regime.

The presence of a hot dark sector in the low-reheat scenario is also consistent with cosmological observations, particularly with the constraints from Big Bang Nucleosynthesis (BBN). In our scenario, the thermal energy density of the DM particles remains suppressed by the negative chemical potential, thereby satisfying the bound on the energy density of additional relativistic species during BBN. We discuss these constraints in detail in the following section, where we identify the corresponding allowed region of the DM parameter space.

\begin{figure}[t]
\hspace{-1 cm}
\includegraphics[scale=0.65]{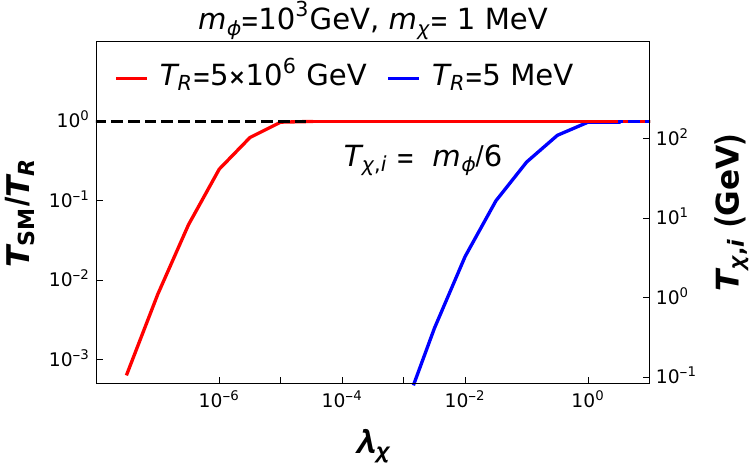}
\includegraphics[scale=0.55]{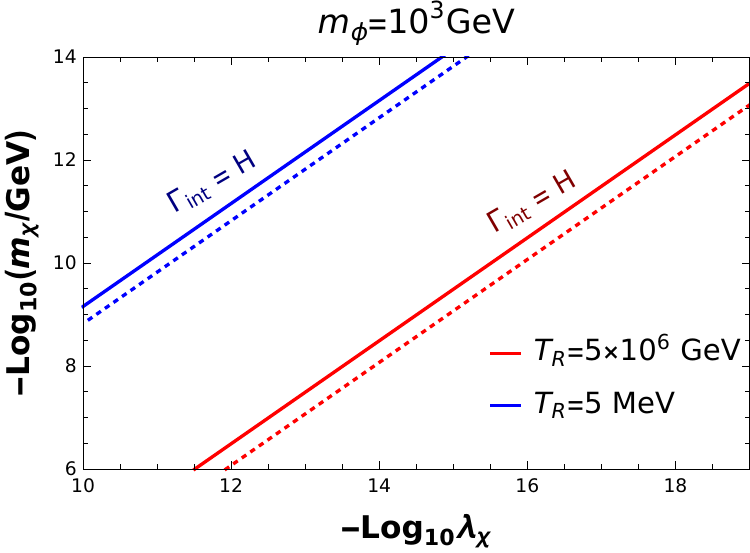}
\caption{\textit{Left panel}: Scaled thermalization timescale ($T_{\rm SM}/T_R$) shown for two benchmark reheat temperatures, $T_R=5\times 10^6$ GeV (red solid line) and $T_R=5$ MeV (blue dot-dashed line ) as a function of quartic self-interaction strength, $\lambda_\chi$. For both cases, $T_{\rm SM}/T_R$ approaches unity as $\lambda_\chi$ exceeds a critical value, indicating the instantaneous thermalization of DM particles. $T^{\rm th}_{\chi,i}$ denotes the thermodynamic temperature just after the thermalization. \textit{Right panel}: DM parameter space for thermalization in the two benchmark scenarios. The right side of each contour line are contours shows DM parameter space where DM particles never thermalize. Solid lines represent the $\Gamma_{\rm int} = H$ contour, derived analytically as in Eq.\ref{eq:notherm}  , whereas dotted lines are found from numerical analysis. See text for details. }
\label{fig:timescale}
\end{figure} 

\subsection{Thermalization timescale}
Although the average quantities, namely $T_\chi, \mu_\chi$, do not reflect the effect of DM self-interaction, thermalization timescale does. The left panel of Fig.\ref{fig:timescale} depicts the time taken for thermalization from the initial epoch, i.e. $T_{SM}=T_R$, as a function of quartic self-interaction strength, $\lambda_\chi$. We show the timescale  in terms of the SM bath temperature scaled by the reheat temperature, fo previous two benchmark scenarios. Thermalization timescale has been obtained via inspecting the numerical distribution function with the MB form at different time-slices. It is evident that thermalization timescale is shorter for stronger self-coupling. In particular, the instantaneous thermalization is achieved for $\lambda_\chi$, larger than some critical value, which is well within the perturbative limit. For instantaneous thermalization, the thermodynamic temperature of DM particle is given by  $T^{\rm th}_{\chi,i} = m_\phi/6$ at the time of DM production, identical to the effective temperature defined for initial non-thermal distribution. 

In the right panel of Fig.\ref{fig:timescale}, we show the DM parameter space in two reheating scenarios as before, for which thermalization can be achieved within the age of universe. This has also been found both numerically (dotted lines) and analytically (solid lines). The right hand side of the dotted lines show the region where DM particles never thermalize. For a complete numerical approach, the evolution of $f_\chi(p,t)$ has been inspected for several combinations of $\lambda_\chi$ and $m_\chi$. We find that it is sufficient to investigate $f_\chi(p,t)$ till DM particles become non-relativistic. This is because elastic scatterings become inefficient in the non-relativistic regime to redistribute DM momenta into a thermal form. This can be understood via comparing the self-interaction rate, $\Gamma_{\rm int}$ with the Hubble expansion rate. In the non-relativistic regime, $\Gamma_{\rm int}/H$ scales as $1/a(t)$ and soon becomes suppressed than unity. Hence, DM thermalization should be complete by the end of the relativistic regime for a suitable choice of $\lambda_\chi$ and $m_\chi$. 

To show it analytically, we first write the self-scattering rate as, 
\begin{align}
\Gamma_{\rm int}(T_{\rm SM}) \simeq n_\chi (T_{\rm SM}) \frac{\lambda^2_\chi}{64\pi E^2 (T_{\rm SM})} = \frac{\Omega_{\chi} \rho_c}{s_0 m_\chi}s(T_{\rm SM})\frac{\lambda^2_\chi}{64\pi  E^2(T_{\rm SM})},
\label{eq:Gamma}
\end{align}
where $E$ is the energy  of each colliding DM particle, Initially, $E=m_\phi/2$ and it redshifts as $1/a(t)$ in the relativistic phase. For non-relativistic DM, the $E$ is dominated by $m_\chi$. The Hubble rate, $H(T_{\rm SM}) = \sqrt{\frac{\pi^2}{90}g_*(T_{\rm SM})}\frac{T^2_{\rm SM}}{M_P}$, which also red-shifts as $1/a(t)$. Therefore, the ratio of the scattering rate to the Hubble rate is given by,
\begin{align}
\frac{\Gamma_{\rm int}}{H}= \left(\frac{\lambda_\chi}{10^{-15}}\right)^2 \left(\frac{2.6\times 10^{-12} ~\rm GeV}{m_\chi}\right)^2 ~\frac{g_{*S}(T_{\rm SM})}{g^{1/2}_*(T_{\rm SM})} 
\frac{T_{\rm SM}}{E(T_{\rm SM})}\frac{m_\chi}{E(T_{\rm SM})}.
\end{align}
It is apparent that the ratio scales as $T_{SM}$, since $E=m_\chi$ in the non-relativistic regime. In the relativistic limit, the ratio grows with the scale factor due to red-shift of $T_{\rm SM}/E^2$. We set the no-thermalization condition as $\frac{\Gamma_{\rm int}}{H} \lesssim 1$, which is imposed at the end of relativistic phase. Notably, we can relate $T_{\rm SM}/E(T_{\rm SM})$ with its initial value in the relativistic regime as the following:
\begin{align}
\frac{T_{\rm SM}}{E(T_{\rm SM})}= \frac{g^{1/3}_{*S}(T_R)}{g^{1/3}_{*S}(T_{\rm SM})}\frac{2 T_R}{m_\phi} .  
\end{align}
Now, combining the above two equations and assuming $m_\chi/E (T_{\rm SM})=1/3$ at the end of relativistic regime, i.e. $T_{\rm SM}=T_{\rm NR}$, we get the following parametric expression between DM mass and self-coupling for no-thermalization condition, $\Gamma_{\rm int}=H$. 
\begin{align}
m_\chi=3.2\times10^{-10}~{\rm GeV}
\!\left(\frac{\lambda_\chi}{10^{-15}}\right)
\!\left(\frac{g_{*S}(T_R)}{100}\right)^{1/6}
\!\left(\frac{T_R}{5\times10^6~{\rm GeV}}\right)^{1/2}
\!\left(\frac{10^3~{\rm GeV}}{m_\phi}\right)^{1/2}
\!\frac{g_{*S}^{1/3}(T_{\rm NR})}{g_*^{1/4}(T_{\rm NR})}
\label{eq:notherm}
\end{align}
The analytical estimate agrees with the numerical results shown in Fig.~\ref{fig:timescale} up to a $10\%$ error  without the undetermined the relativistic degrees of freedom at $T_{\rm SM}=T_{\rm NR}$. In fact, $T_{\rm NR}$ can only be determined only with numerical analysis, comparing the average energy of DM particles with their mass.

\section{Cosmological constraints}
\label{sec:sec3}
In the previous section, we have shown the evolution of DM temperature and chemical potential for relic satisfied scenarios by fixing $\beta$, while $T_R, m_\phi$ and $m_\chi$ remain free parameters. In addition, in presence of elastic scattering the DM temperature, representative of average kinetic energy, is completely fixed by initial conditions, irrespective of its thermalization history. Therefore, the self-coupling becomes completely unconstrained, although it facilitates the thermalization process. In this section, we are going to discuss cosmological and astrophysical observations which can constrain $m_\chi$ and $\lambda_\chi$ for both high and low reheat scenarios, as introduced before.  

\subsection{Constraints from BBN, CMB and Bullet Clusters}

The observation of primordial light element abundances is aligned with the predictions of standard BBN calculations in the radiation-dominated universe to a great accuracy. Therefore, the number of extra particle species, significantly contributing to the Hubble rate other than the SM radiation is constrained. This is parametrized in terms of the effective number of extra neutrino species ($\Delta N_{\rm eff}$) as follows : 
\begin{align}
\Delta N_{\rm eff} = \frac{8}{7} \frac{\rho_R}{\rho_\gamma},
\end{align}  
where $\rho_R$ is total energy density of extra relativistic particles and the $\rho_\gamma$ is the photon counterpart. This definition of $\Delta N_{\rm eff}$ is valid before the freeze-out of electron-positron annihilation.  The number of effective neutrino species during BBN is $N_{\rm eff} =2.86 \pm 0.15$ \cite{Fields:2019pfx}, whereas $N^{\rm SM}_{\rm eff} \simeq 3$ is expected within SM. This fixes the central value of $\Delta N_{\rm eff}$ to be $0.14$. 

For our case, we can put the upper limit on the DM density at the BBN epoch, $T_{\rm SM} \sim 1 $ MeV in our present scenario. The energy density of the secluded sector DM  can be written for both relativistic and non-relativistic regimes as,
\begin{align}
\rho_\chi (T_{\rm SM}) =  
\begin{cases}
  3 T^{\rm eff}_\chi(T_{\rm SM}) n_\chi (T_{\rm SM})~~ \text{for $m_\chi/T_\chi\lesssim 3$},\vspace{0.2cm}\\
   m_\chi n_\chi (T_{\rm SM}) ~ ~\text{for $m_\chi/T_\chi\gtrsim 3$},
\end{cases}
\end{align}
where we have used the definition of $T^{\rm eff}_\chi$ as in Eq.\ref{eq:temp}. As discussed earlier, this expression holds true for both thermal and non-thermal distribution in presence of number-conserving processes. Now we calculate the DM energy density at the onset of BBN ($T_{\rm SM}=T_{\rm BBN}=1$ MeV) replacing $n_\chi (T_{\rm BBN})$ with $n_\chi (T_R)$, normalized to saturate the observed the DM relic density, as the following :
\begin{align}
\rho_\chi (T_{\rm BBN}) =  
\begin{cases}
  6.2\times 10^{-18}~ {\rm GeV^4} \left(\frac{T^{\rm eff}_\chi(T_{\rm BBN})}{m_\chi}\right)~~ \text{for $m_\chi/T_\chi\lesssim 3$},\vspace{0.2cm}\\
   2.1\times 10^{-18}~ {\rm GeV^4} ~ ~\text{for $m_\chi/T^{\rm eff}_\chi\gtrsim 3$}.
\end{cases}
\end{align}

The observational constraint on $\rho_\chi$ from the estimate of $\Delta N_{\rm eff}$ turns out to be,  $\rho_\chi (T_{\rm BBN}) \lesssim 8.06\times 10^{-14}~{\rm GeV^4}$, which translates to :
\begin{align}
\frac{T^{\rm eff}_\chi(T_{\rm BBN})}{m_\chi} \lesssim 1.3\times 10^4.
\label{eq:bbn}
\end{align}
Note that, for non-relativistic DM, the bound is trivially satisfied. It is apparent that DM particles can be highly relativistic at the time of BBN without violating constraint on $\Delta N_{\rm eff}$. At prime facie, this may appear problematic for a thermal DM scenario, since the thermal energy density of a relativistic species becomes greater than that of SM radiation. Therefore, in most secluded dark matter scenarios \cite{Adshead:2016xxj,Ghosh:2022asg,Ghosh:2023mnh}, the dark sector temperature is typically lower than that of the SM thermal bath, thereby ensuring a suppressed energy density. However, this is only true with a vanishing chemical potential. 

In our scenario, the DM energy density becomes suppressed as the chemical potential can take negative values in low reheat scenario. It is apparent from Eq.\ref{eq:chem} the chemical potential becomes negative at the time of BBN for $\frac{T^{\rm eff}_\chi(T_{\rm BBN})}{T_{\rm BBN}}>\left(\frac{2\times 10^{-8}{~\rm GeV}}{m_\chi}\right)^{1/3}$, where
\begin{align}
 \frac{T^{\rm eff}_\chi (T_{\rm BBN)})}{T_{\rm BBN}} = \frac{m_\phi}{6 T_R}\left(\frac{10}{g_{*S}(T_R)}\right)^{1/3} 
\label{eq:tempexp}
\end{align}
 We note that $T^{\rm eff}_\chi (T_{\rm BBN}) > T_{\rm BBN}$ for $m_\phi \gtrsim 6 T_R$ with mild variation of effective degrees of freedom ($g_{*S}$) related to the entropy density of the universe. This is perfectly allowed by the BBN constraint shown in Eq.\ref{eq:bbn}, as long as $m_\chi \gtrsim 10^{-7}$ GeV. This stems from the fact that the constant inflaton mass fixes the initial momentum average of DM particles, while the duration between the reheating and the BBN determines the red-shift of DM momenta with effects of elastic scatterings. In particular, for low reheat temperature, i.e. $T_R= 5$ MeV, there is little time to reduce the momentum average, therefore, a hotter DM bath emerges during BBN. Combining Eq.\ref{eq:tempexp} and Eq.\ref{eq:bbn}, we find the lower limit on the DM mass from BBN
\begin{align}
m_\chi \gtrsim 7.7\times 10^{-8} {~\rm GeV} \left(\frac{m_\phi}{6T_R}\right)\left(\frac{10}{g_{*S}(T_R)}\right)^{1/3},
\label{eq:mbbn}
\end{align}
independent of $\lambda_\chi$, which completely hides thermalization history. To reiterate, the BBN bound only depends on the energy density, not on the shape of the DM distribution function. For our case, the energy density is already fixed by initial conditions.

Now, we shall look into constraints from the CMB background power spectrum, which demands DM to be cold enough at the time of last scattering of photons, i.e. $T_{\rm CMB}\approx 0.26$ eV. In particular, the CMB data can impose constraints on secluded DM scenario as the following \cite{Heimersheim:2020aoc,Ghosh:2022asg} : 
\begin{align}
\frac{T^{\rm eff}_{\chi} (T_{\rm CMB})}{m_\chi} < 10^{-5},
\end{align}
which implies a highly non-relativistic DM at the time of last scattering of photons. Following similar approach as before, we can derive a lower bound on the DM mass analytically in terms of initial parameters. However, this time, we need to take care of both relativistic and non-relativistic evolution of DM temperature from the initial epoch ($a_{\rm in}$) to the CMB epoch ($a_{\rm CMB}$) with a hierarchy, $a_{\rm in} < a_{\rm NR} < a_{\rm CMB}$, where $a_{\rm NR}$ corresponds to the onset of non-relativistic regime, i.e. $m_\chi/T^{\rm eff}_{\chi}\approx 3$. Therefore, the DM temperature at the time of CMB can be written as
\begin{align}
T^{\rm eff}_\chi(T_{\rm CMB}) = \frac{T^{\rm eff}_\chi(a_{\rm in})^2}{T^{\rm eff}_\chi(a_{\rm NR}) } \left(\frac{a_{\rm in}}{a_{\rm CMB}}\right)^2 = \frac{m^2_\phi}{12 m_\chi}\left(\frac{T_{\rm CMB}}{T_R}\right)^2 \left(\frac{g_{*S}(T_{\rm CMB})}{g_{*S}(T_R)}\right)^{2/3},
\end{align} 
where we have put $T_\chi(a_{\rm in})= m_\phi/6$ and $T_\chi(a_{\rm NR})= m_\chi/3$. Consequently, the lower bound on the DM mass from the CMB observations is found to be
\begin{align}
m_\chi \gtrsim 1.4\times 10^{-7} {~\rm GeV} \left(\frac{m_\phi}{6 T_R}\right)\left(\frac{3.91}{g_{*S}(T_R)}\right)^{1/3}.
\end{align}
To note, the lower bound on the DM mass derived from the CMB constraint is stronger by a $32\%$ than that of the BBN, shown in Eq.\ref{eq:mbbn}. This is in contrast with canonical thermal DM scenarios (thermalized with the SM bath), where the BBN constraint put stronger lower bound on the DM mass, i.e. around a MeV \cite{Nollett:2013pwa,Boehm:2013jpa,Ghosh:2022frt}, whereas the observation of large scale structure with the CMB data restricts thermal DM scenarios with masses approximately less than a few keV \cite{Viel:2005qj,Irsic:2017ixq,Chatterjee:2023mlh}. This stems from the fact that DM, thermalized with the SM thermal bath, can not be relativistic during BBN, i.e. $T_{\rm SM}\sim 1$ MeV. As explained earlier, for a secluded DM with non-zero chemical potential, the BBN constrained can be satisfied with a highly relativistic DM particles.

\begin{figure}[t]
\centering
\includegraphics[scale=0.44]{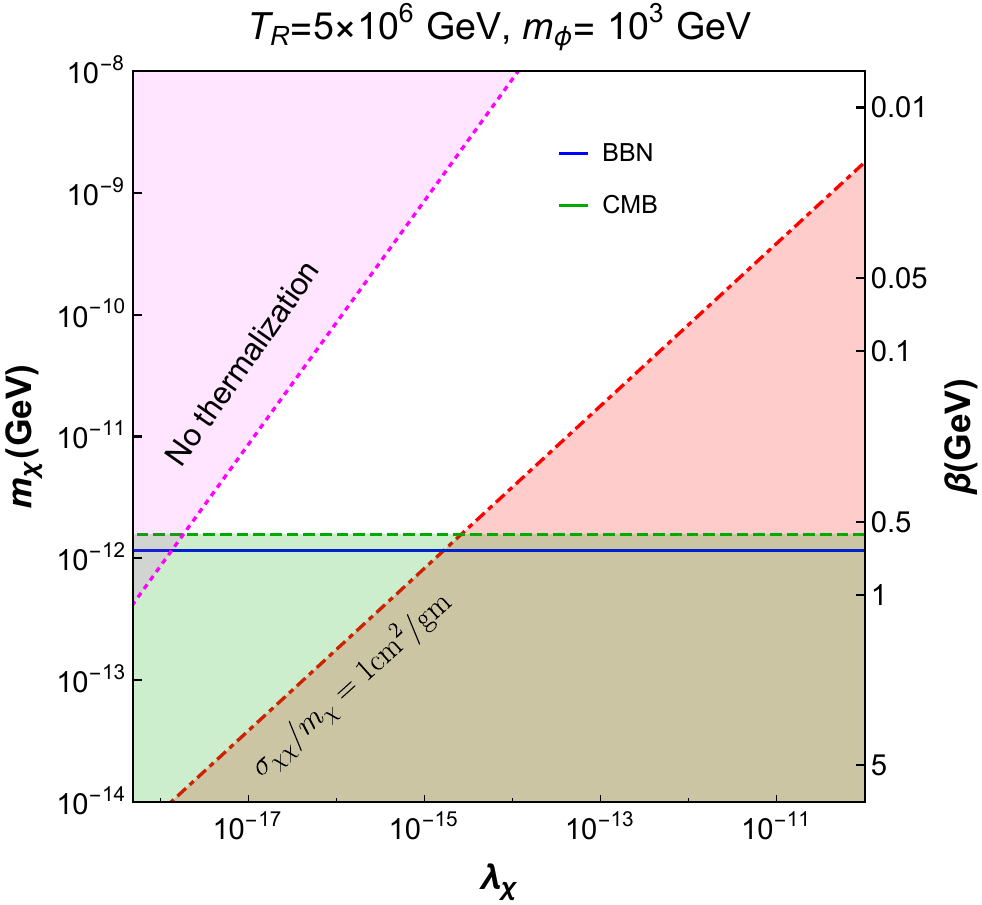}
\includegraphics[scale=0.47]{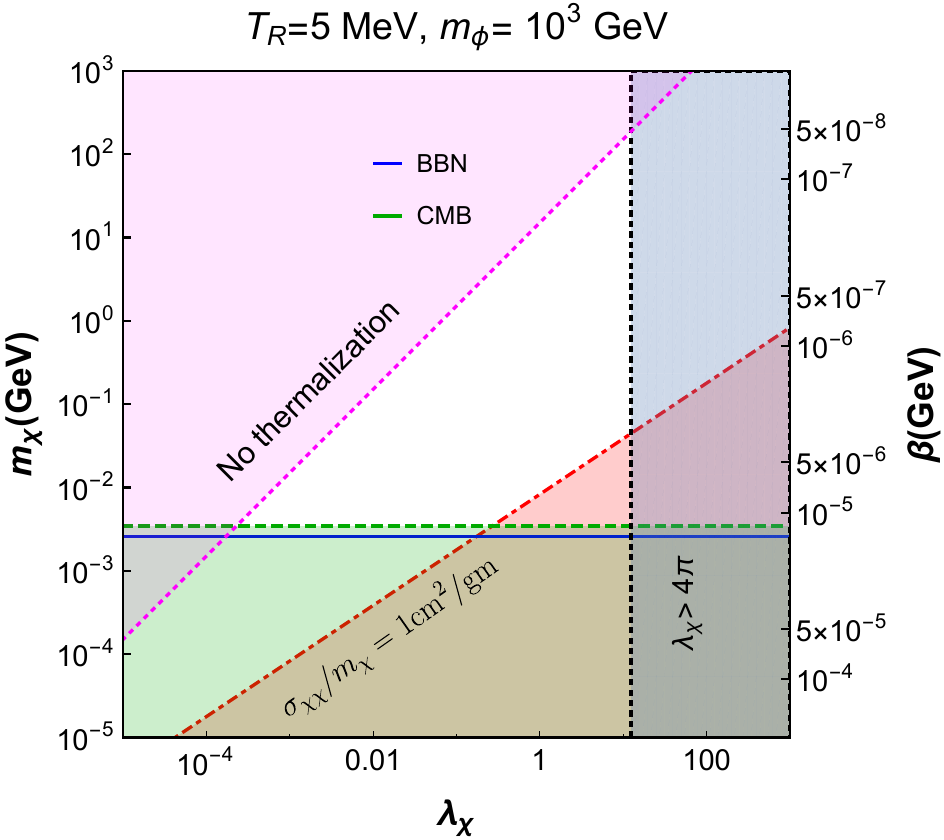}
\caption{\textit{Cosmological constraints on  self-interacting secluded DM} : Depiction of allowed parameter space of the DM mass ($m_\chi$) and its quartic self-coupling ($\lambda_\chi$) for two benchmark values of the reheat temperature, i.e $T_R =5\times 10^6$ GeV (\textit{left}) and $T_R=5$ MeV (\textit{right}), while the inflaton mass is fixed at $m_\phi = 10^3$ GeV. The lower bounds on the DM mass coming from BBN (blue solid line) and CMB (green dashed line) observations have been shown. The lower limit on the DM mass  coming from the Bullet cluster bound is shown by red dot-dashed line. The magenta shaded region shows where DM never thermalizes and the black dotted line in the right panel indicates the perturbative limit of $\lambda_\chi$. }
\label{fig:cosmo}
\end{figure}

The DM mass bounds shown in the above, are independent of DM self-scattering cross-section. However, the self-scattering cross-section is constrained from the merging of Bullet clusters as $\sigma_{\chi\chi}/m_\chi \lesssim 1~ {\rm cm^2/gm}$ \cite{Randall:2008ppe,Robertson:2016xjh}. Within our model, this bound translates to :
\begin{align}
m_\chi \gtrsim 44 ~{\rm MeV}\left(\frac{\lambda_\chi}{4\pi}\right)^{2/3},
\label{eq:bullet}
\end{align}
where we have used $\sigma_{\chi \chi}=\frac{\lambda^2_\chi}{64\pi m^2_\chi}$ for non-relativistic DM particles at late times. This is valid for perturbative values of dimensionless coupling, $\lambda_\chi$. In Fig.\ref{fig:cosmo} we show constraints in the $m_\chi-\lambda_\chi$ plane for two reheat temperatures using limits from BBN, CMB and Bullet clusters. 

In the left panel of Fig.\ref{fig:cosmo}, we depict the case of high reheat temperature, i.e. $T_R=5\times 10^6$ GeV. To remind, all parameter points shown in both figures satisfy the relic density constraint from Planck 2018 Observations \cite{Planck:2018vyg}. Therefore, we show $\beta$ values corresponding to the DM mass using Eq.\ref{eq:relic}, for which the DM relic density is set at the reheating. The green solid line and blue solid line indicate the lower bound on the DM mass from CMB and BBN observations respectively. 
The red dot-dashed line shows the Bullet cluster bound on the DM mass, as derived in Eq.\ref{eq:bullet}. This bound becomes stronger for $\lambda_\chi \gtrsim 10^{-15}$ than the CMB bound. Since the merging of galaxies is a late-time phenomena, the DM velocity dispersion is not decided by the DM temperature, rather by the gravitational potential of galaxies. Therefore, the Bullet cluster bound is only dependent on the DM self-scattering cross-section and its mass. The white region shows the cosmologically allowed parameter space of a thermal secluded DM. The magenta-shaded region above the green dashed line is also allowed from cosmologically observations, although in that region DM momentum distribution remains non-thermal. The observational difference between thermal and non-thermal distribution may be seen in matter power spectrum, which requires a dedicated study in cosmological perturbation theory, thereby left for future work.

In the right panel, same constraints are shown for $T_R = 5 $ MeV. Unlike the high reheat temperature scenario, the DM temperature is higher than the SM one at the initial epoch, since $m_\phi> 6T_R$. Then the DM momentum red-shift happens for a shorter time period, starting from the initial epoch up to the BBN/CMB epochs, which keeps the dark sector hotter than the former case. In addition, the DM number density produced at the initial epoch is smaller for lower reheat temperature, evident from Eq.\ref{eq:num}. Therefore, a larger DM mass is required to satisfy the relevant cosmological observations. In particular, the lower bound on the DM mass from the CMB is approximately 9 orders of magnitude stronger than that of the high-reheat scenario. For thermalization, a stronger $\lambda_\chi$ is required to attain the necessary elastic scattering rate, which is otherwise suppressed for larger DM masses. As a result, the parameter space of a cosmologically allowed secluded thermalized DM is reduced compared to the high reheat scenario, as apparent from the white regions above the green dashed line in Fig.\ref{fig:cosmo}.

As apparent from Fig.\ref{fig:cosmo}, the lower limit on the mass of a secluded DM scenario can be quite different from warm DM scenarios \cite{Viel:2005qj,Irsic:2017ixq,Chatterjee:2023mlh}. For high reheat scenario, $m_\chi \gtrsim 10^{-12}$ GeV, which is 6 orders of magnitude smaller than the oft-quoted keV bound in the warm DM case. On the other hand, $m_\chi \gtrsim 10^{-3}$ GeV for low reheat scenario with the same choice of inflaton mass. Therefore, for a self-interacting secluded DM scenario, the inflationary and post-inflationary physics become the deciding factors in determining DM parameters.

\subsection{Constraints from the free-streaming length}
In cases of BBN and CMB observations, the DM mass bound is controlled by the effective DM temperature and the chemical potential, which are set by initial conditions. Since there is only elastic scattering present, these bounds are found to be independent of thermalization process or effects of collisions in the early universe. The effects of collision becomes apparent in the DM free-streaming length calculation. As we know, the DM perturbations become Jeans unstable during the matter-radiation equality epoch ($t=t_{\rm eq}$). Hence, depending on the DM velocity, DM particles can stream out from a over-dense region to under-dense region, thereby suppressing the growth of structures. This effect can approximately be understood via co-moving free-streaming length ($\lambda_{\rm fsl}$) of DM particles, which is defined as follows \cite{Kolb:1990vq} :
\begin{align}
    \lambda_{\rm fsl} = \int^{^{t_{\rm eq}}}_{t_{\rm dec}} \frac{\expval{v(t)}}{a(t)} dt ,
\end{align}
where $t_{\rm dec}$ denotes the time when elastic collisions stop and DM particles no longer behaves like an interacting fluid. For the non-thermal case, the elastic collisions are not efficient. Hence, we can consider DM particles collisionless since their production, i.e. $t_{\rm dec}=t (T_R)$. The average velocity, $\langle v(t)\rangle$, is  defined as
\begin{align}
   \expval{v(t)} = \frac{\int\frac{d^3 p}{(2\pi)^3} \frac{p}{\sqrt{p^2+m^2_\chi}}f_\chi (p,t)}{\int\frac{d^3 p}{(2\pi)^3} f_\chi (p,t)}.
   \label{eq:vel}
\end{align}
For the non-thermal (collisionless) case,  $\expval{v(t)}$ can be evaluated in terms of initial conditions using $f_{\chi}(p,t)= A \delta \left(p-\frac{m_\phi}{2} \frac{a(T_R)}{a(T)}\right)$,  as follows :
\begin{align}
     \expval{v(a)} = \frac{\frac{m_\phi}{2}\frac{a(T_R)}{a(T)}}{\sqrt{\left(\frac{m_\phi}{2}\frac{a(T_R)}{a(T)}\right)^2+m^2_\chi}},
\end{align}
 where $a(T_R)= \left(\frac{g_{*S}(T_{\rm eq})}{g_{*S}(T_R)}\right)^{1/3}\left(\frac{T_{\rm eq}}{T_R}\right) a(T_{\rm eq})$. To note, for $ m_\chi \ll \frac{m_\phi}{2}\frac{a(T_R)}{a(T)}$, DM particles are relativistic, hence $\expval{v(a)}=1$. In this regime, the contribution to the  co-moving free-streaming length is independent of DM mass, whereas for non-relativistic regime, the contribution will be inversely proportional to the DM mass. 
 
 We now express $\lambda_{\rm fsl}$ in terms of the scale factor, using $ H (a) = \frac{da}{adt}= H_0 a(T)^{-2}\sqrt{\Omega_{R,0}}$ $\sqrt{1+a(T)/a(T_{\rm eq})} $, where $H_0$ is the Hubble parameter and $\Omega_{R,0}$ radiation density parameter in the current universe 
 \begin{align}
     \lambda_{\rm fsl} = \int^{a(T_{\rm eq})}_{a(T_R)} \frac{\expval{v(a)} d a}{H_0\sqrt{\Omega_{R,0}}\sqrt{1+a(T)/a(T_{\rm eq})}},
 \end{align}
 where the scale factor at the time of matter-radiation equality epoch is given by, $a(T_{\rm eq})=2.91\times 10^{-4}$ \cite{Planck:2018vyg}. For collisional case, we calculate Eq.\ref{eq:vel} using the DM momentum distribution function found numerically solving the Boltzmann equation, followed by the numerical evaluation of the above equation, replacing $a(T_R)$ by $a(T_{\rm dec})$, where $a(T_{\rm dec})$ denotes the epoch when the elastic scatterings stop due to the Hubble expansion.  
\begin{figure}[t]
\centering
\includegraphics[scale=0.30]{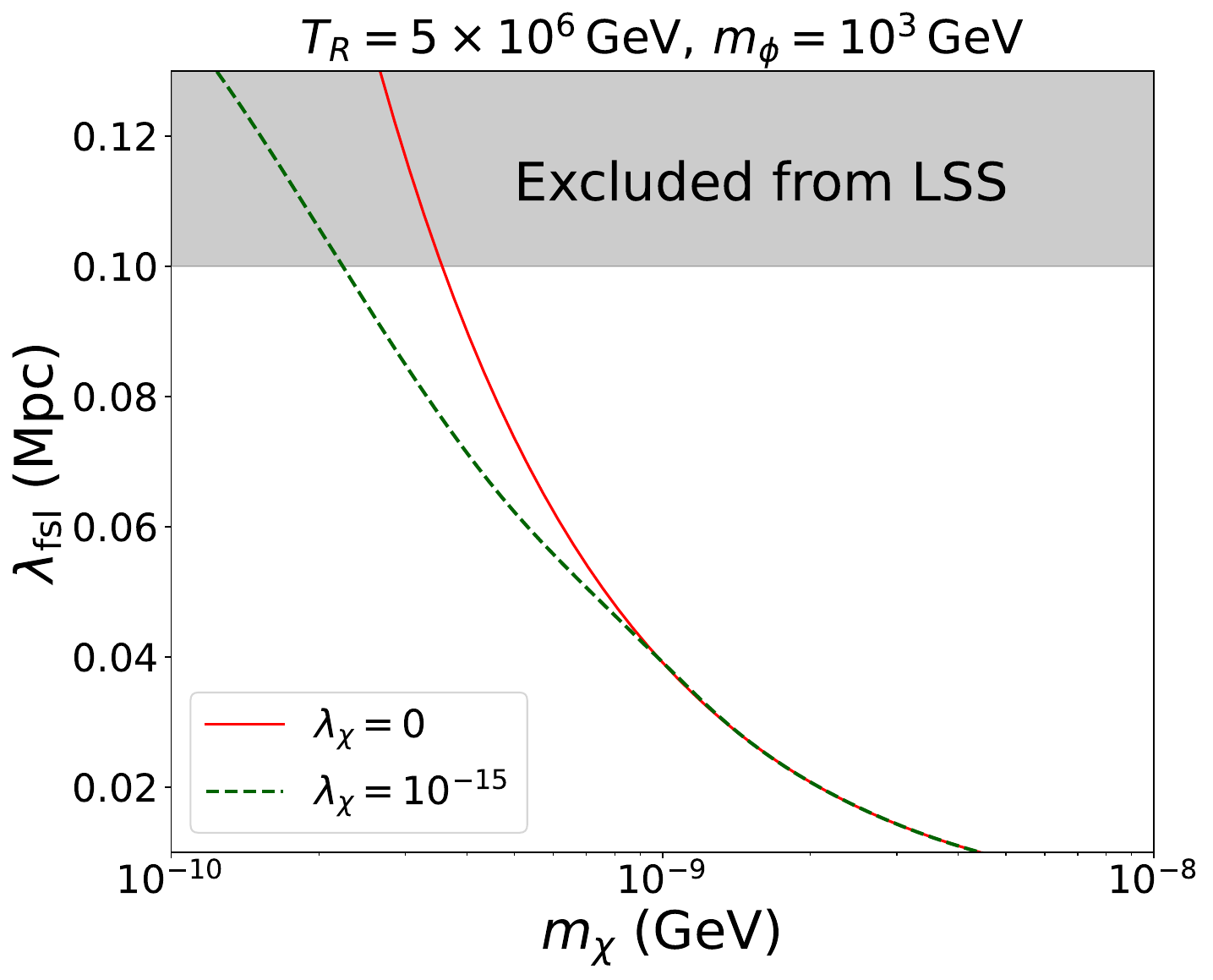}
\includegraphics[scale=0.30]{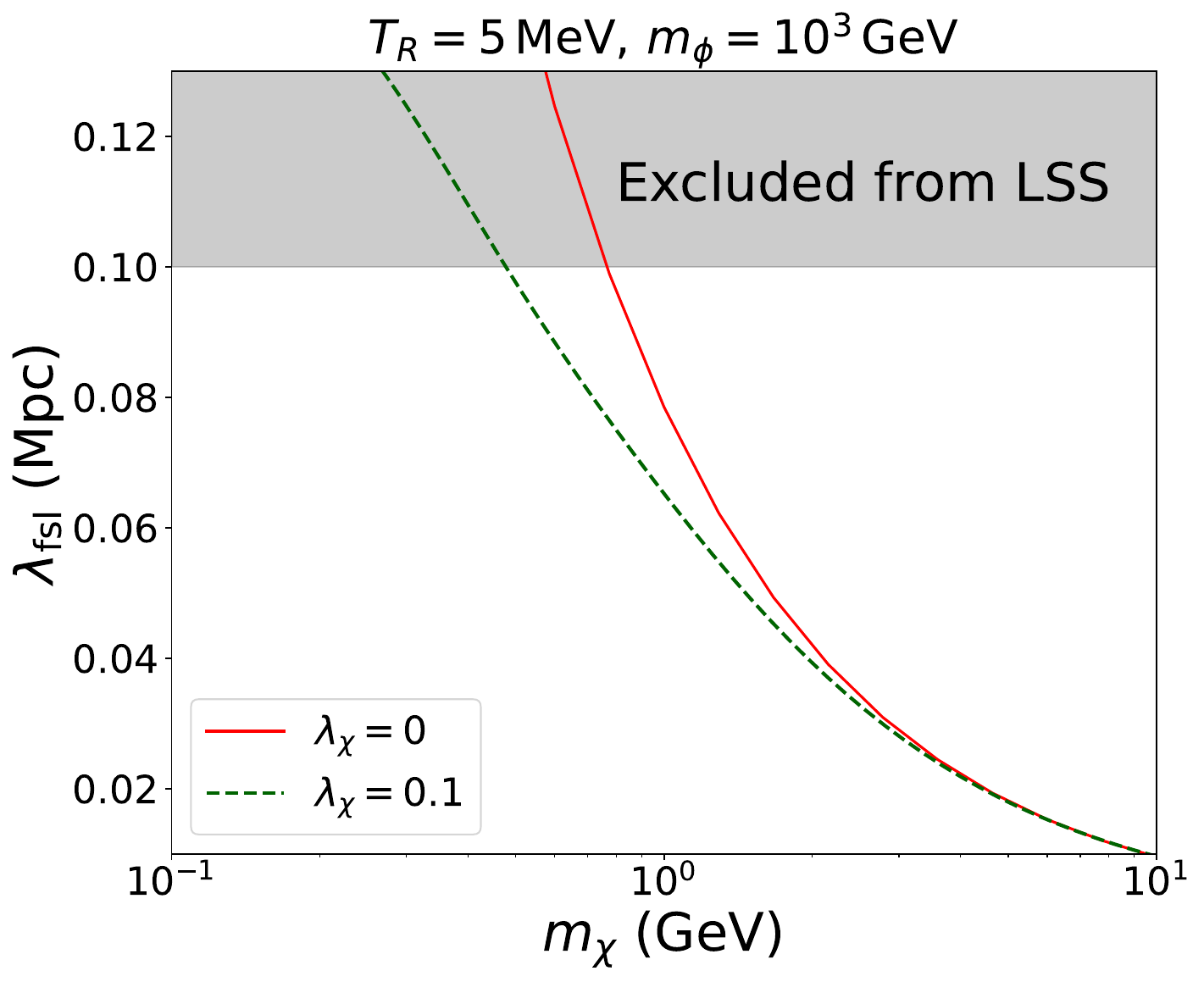}
\caption{The co-moving free-streaming length ($\lambda_{\rm fsl}$) of DM particles shown for non-thermal (collisionless) and thermal (collisional) distributions as a function of DM mass. $\lambda_{\rm fsl} >0.1$ Mpc is excluded from large scale structure (LSS) formation, especially from the existence of DM-rich dwarf galaxies of size around 1 kpc.}
\label{fig:fsl}
\end{figure}

In Fig.\ref{fig:fsl}, we depict $\lambda_{\rm fsl}$ for two reheat scenarios and compare the thermal and non-thermal cases. The non-thermal case is represented by $\lambda_\chi=0$ scenario and shown in solid red lines in both panels of Fig.\ref{fig:fsl}, whereas green-dashed lines indicate the collisional scenario with non-zero values of $\lambda_\chi$. For collisional scenario, the free-streaming of DM particles is expected to be delayed compared to the collisionless scenario, depending on the strength of self-scattering cross-section determined by $m_\chi$ and $\lambda_\chi$. The decoupling time of elastic scatterings is computed numerically by checking $\Gamma_{\rm int}/H$ throughout the evolution of DM momentum distribution function. In particular, we track the transition from $\Gamma_{\rm int}/H>1$ to $\Gamma_{\rm int}/H < 1$ to get the decoupling time for given values of $m_\chi$ and $\lambda_\chi$. For a graphical representation please see Fig.\ref{fig:GammaByH} in Appendix \ref{app:B}. For collisionless scenario, $\Gamma_{\rm int}/H < 1$ throughout the cosmic history. 

For both scenarios, we show $\lambda_{\rm fsl}$ for a DM mass range for which $\lambda_{\rm fsl}$ varies as $1/m_\chi$, indicating the non-relativistic contribution as mentioned before. Such a choice is apparent as structure formation is suppressed for relativistic particles with large free-streaming length. In the left panel, the collisionless  and collisional scenarios coincide for $m_\chi>10^{-9}$ GeV, since the elastic scatterings become inefficient for $\lambda_\chi = 10^{-15}$. Similar feature is seen for low reheat scenario in the right panel for $m_\chi> 5$ GeV and $\lambda_\chi=0.1$. The difference in high and low reheat scenario stems from the duration of momentum red-shift between the production time and the matter-radiation equality epoch. For lower DM masses, the collisional effect becomes prominent due to enhanced elastic scattering cross-section. Consequently, the co-moving free-streaming length in the collisional scenario becomes smaller compared to the collisionless scenario. Furthermore, the difference between these two cases increases as the DM mass decreases.     

We also depict the constraint on DM mass from the large scale structure (LSS), considering a typical size of dwarf galaxies, i.e. around 1 kpc. For $\lambda_{\rm fsl} \lesssim 0.1$ Mpc, the dwarf galaxies can form, therefore, we use this to put an lower limit on the DM mass \cite{Rubakov:2017xzr}. The lower limit on the DM mass becomes $3.5\times 10^{-10}$ GeV for high reheat case and $0.77$ GeV for low reheat case in the collisionless limit. With collisional effect, the lower limit weakens approximately by $40\%$ for our choice of self-interaction coupling.  The lower limit on the DM mass differs by several orders of magnitude for two reheat scenarios due to difference in the average velocity, For high reheat scenario, for a given mass the dark sector becomes lot cooler than the low reheat scenario, since the initial average momentum set by the inflaton mass, is identical for both cases.

In the both panels of Fig.\ref{fig:fsl}, we have shown the lower limit on the DM mass for a specific choices of $\lambda_\chi$. Given a DM mass, the decoupling happens earlier for smaller $\lambda_\chi$, thereby coinciding with the collisionless scenario. On the other hand, for higher $\lambda_\chi$, the dark sector remains a collisional fluid till the matter-radiation equality epoch, rendering a negligible free-streaming length. Therefore, we suitably choose $\lambda_\chi$ to demonstrate the effect of collision in determining the lower limit of the DM mass using LSS constraints.     

To summarize, for a self-interacting secluded DM, the constraint on the mass is heavily dependent on its initial production mechanisms, especially on the inflaton mass and the reheat temperature. The BBN and CMB constraints are independent of thermalization process as long as DM particles are participating in only elastic scatterings. The effect of elastic collisions, in turn the process of thermalization becomes apparent cosmologically via the free-streaming length which is suppressed due to collision. In the late universe, we have a weak bound on this collisions from merging of bullet clusters, which is complementary to these constraints from the early universe cosmology.

\section{Conclusion $\&$ Future Outlook}
\label{sec:sec4}
 In this work, we have studied the evolution of the momentum distribution function of a secluded self-interacting scalar DM, whose initial condition is set by the inflaton decay at the time of reheating. We have demonstrated how $2\rightarrow 2$ elastic scattering mediated by the quartic self-coupling drives the distribution toward the Maxwell--Boltzmann form for a given inflaton mass and reheat temperature, and we have followed the subsequent evolution of the DM temperature and chemical potential. For sufficiently large quartic self-coupling, the initial DM temperature is set by the inflaton mass as $T^i_{\rm DM}=m_\phi/6$, while the chemical potential, fixed by the conserved DM number, can be positive or negative depending on $m_\phi$ and $T_R$. In the presence of a non-zero chemical potential, the DM temperature can be larger than the visible one while satisfying all BBN and CMB constraints. We believe this is an important novel result, since in most secluded DM scenarios the DM temperature turns out to be smaller than the visible one.

The lower limits on the DM mass from BBN and CMB, $\sim 10^{-12}$~GeV and $\sim 10^{-3}$~GeV for $T_R=5\times 10^6$~GeV and $5$~MeV respectively, are independent of the self-interaction strength, since the DM temperature is set by the initial condition rather than by the details of thermalization. The CMB bound turns out to be about $32\%$ stronger than the BBN one, in contrast to the canonical thermal DM case, where BBN yields the stronger bound. The effect of collisions becomes apparent only in the co-moving free-streaming length, where elastic scatterings weaken the lower bound on the DM mass by about $40\%$ and introduce an explicit dependence on the self-interaction strength. We obtain lower limits on the thermal DM mass of $2.2\times10^{-10}$~GeV and $0.47$~GeV for $T_R=5\times 10^6$~GeV and $5$~MeV respectively, with $m_\phi=10^3$~GeV. Therefore, the post-inflationary history and the poorly constrained DM self-interaction are the deciding factors in constraining secluded DM with a non-zero chemical potential, and the predictability of the lower mass limit weakens due to its dependence on initial conditions.

We have performed the current study with simplified assumptions of instantaneous and perturbative decay of the inflaton and followed by the onset of the radiation dominated universe, which led to an sharply peaked initial momentum distribution function for DM particles. We expect a broadening of this initial distribution function if these assumptions are relaxations. In particular, the initial distribution function from the non-perturbative decay of inflaton is expected to be quite different than the present one, thereby altering the thermalization history. Moreover, the presence of number-changing process within the dark sector and with the visible sector can alter the thermodynamic properties, thereby affecting the constraints on DM parameter space. This warrants a dedicated future study including non-trivial effects of the inflaton decay and all possible scattering processes of DM particles \cite{future}.  

\section*{Acknowledgement}
DG acknowledges the Post-doctoral Fellowship from SINP Kolkata. AK acknowledges the Senior Research Fellowship (SRF) awarded by the University Grants Commission (UGC), Government of India.

\appendix
\section{Details of the Boltzmann Equation}
\label{app:boltz}
\subsection{Boltzmann Equation in an FLRW Universe} 
\label{app:gen}

For a general $2 \rightarrow 2$ scattering process such as
$A(p_1) + B(p_2) \leftrightarrow C(p_3) + D(p_4)$, the Boltzmann equation
governing the evolution of the phase-space distribution function of particle $A$
with momentum $p_1$ reads~\cite{Kolb:1990vq, Dodelson:2003ft}
\begin{equation}
    \begin{aligned}
        \frac{d f_A(\vec{p}_1,\vec{r},t)}{dt}
        = & -\frac{1}{2E_1} \int \frac{d^3 p_2}{(2\pi)^3\, 2E_2}
        \int \frac{d^3 p_3}{(2\pi)^3\, 2E_3}
        \int \frac{d^3 p_4}{(2\pi)^3\, 2E_4} \\
        & \times (2\pi)^4\, \delta^4\!\left(p_1+p_2-p_3-p_4\right)\,
        |\mathcal{M}|^2 \left[f_A(p_1) f_B(p_2) - f_C(p_3) f_D(p_4)\right],
    \end{aligned}
    \label{gen_eq}
\end{equation}
where $|\mathcal{M}|^2$ is the squared matrix element for the given process,
summed over the internal degrees of freedom of the initial and final states.
Bose enhancement and Pauli blocking term are neglected for simplicity.

Incorporating quantum statistics leaves the phase-space integrals, the delta function and $|\mathcal{M}|^2$ in Eq.~\eqref{gen_eq} unchanged; only the bracketed combination of distribution functions is modified to $f_A(p_1) f_B(p_2)\,[1\pm f_C(p_3)][1\pm f_D(p_4)] - f_C(p_3) f_D(p_4)\,[1\pm f_A(p_1)][1\pm f_B(p_2)]$, where the upper (lower) sign applies to bosons (fermions), accounting for Bose enhancement (Pauli blocking) of the final states~\cite{Uehling:1934zz}. The collision term thereby becomes nonlinear in the distribution functions, and the equilibrium solution is the Bose--Einstein (Fermi--Dirac) distribution instead of the Maxwell--Boltzmann one. In the dilute limit, $f\ll 1$, the additional factors reduce to unity and Eq.~\eqref{gen_eq} is recovered.

In an FLRW Universe, where the spatial part is assumed to be homogeneous and
isotropic, the distribution function does not depend on $\vec{r}$, and the
free-streaming momentum redshifts as $\dot{p} = -Hp$. In this case,
Eq.~\ref{gen_eq} can be written as
\begin{equation}
\begin{aligned}
\frac{\partial f_A(p_1,t)}{\partial t}
-H|\vec p_1|\frac{\partial f_A(p_1,t)}{\partial |\vec p_1|}
=-\frac{1}{2E_1}\int d\Pi_2\,d\Pi_3 \, &d\Pi_4 \, (2\pi)^4\delta^4(p_1+p_2-p_3-p_4)
\,|\mathcal M|^2 \\
&\times \left[f_A(p_1)f_B(p_2)-f_C(p_3)f_D(p_4)\right],
\end{aligned}
\label{app:f_eq_phy}
\end{equation} where $H$ is the Hubble parameter and as shorthand, we have used $d\Pi_i = \dfrac{d^3p_i}{(2\pi)^3\, 2E_i}$.

To obtain the Boltzmann equation at the number-density level, one integrates
Eq.~\ref{app:f_eq_phy} over $\vec{p}_1$ (with measure $d^3p_1/(2\pi)^3$).
By doing so, the evolution equation for the number density
$n_A=\int \frac{d^3p}{(2\pi)^3}\, f_A$ can be written as
\begin{equation}
    \frac{d n_A}{dt} + 3Hn_A
    = -\int d\Pi_1\, d\Pi_2\, d\Pi_3\, d\Pi_4\,
    (2\pi)^4\, \delta^4\!\left(p_1+p_2-p_3-p_4\right)\,
    |\mathcal{M}|^2 \left[f_A f_B - f_C f_D\right].
    \label{app:n_eq_phy}
\end{equation}  

Similarly, the Boltzmann equation for the energy density
$\rho_A=\int\frac{d^3p}{(2\pi)^3}\,E_A f_A$ is
\begin{equation}
    \frac{d\rho_A}{dt}+3H(\rho_A+P_A)
    = -\int d\Pi_1\, d\Pi_2\, d\Pi_3\, d\Pi_4\, E_A(p_1)\,
    (2\pi)^4\delta^4(p_1+p_2-p_3-p_4)\,
    |\mathcal{M}|^2\left[f_Af_B-f_Cf_D\right],
    \label{app:rho_eq}
\end{equation}
where $P_A=\int\frac{d^3p}{(2\pi)^3}\frac{p^2}{3E_A}f_A$ is the pressure. In the
relativistic regime $P_A=\rho_A/3$, so the left-hand side becomes
$\dot\rho_A+4H\rho_A$, while in the non-relativistic regime $P_A\ll\rho_A$ and it
becomes $\dot\rho_A+3H\rho_A$.

For elastic scattering of identical particles, the right-hand sides of
Eqs.~\ref{app:n_eq_phy} and \ref{app:rho_eq} vanish, as shown in
Sec.~\ref{app:final_form_BE} (the latter due to energy conservation in each
collision). Our main goal is therefore to check that the number density and the energy density, computed by integrating the numerical solution $f_A(p,t)$ of Eq.~\ref{app:f_eq_phy}, satisfy Eqs.~\ref{app:n_eq_phy} and
\ref{app:rho_eq}, i.e.\ that
\begin{equation}
    n_A \propto \frac{1}{a^3}, \qquad
    \rho_A \propto
    \begin{cases}
        1/a^4 & \text{(relativistic regime)},\\
        1/a^3 & \text{(non-relativistic regime)}.
    \end{cases}
\end{equation}\\

\subsection{Details of the Collision Term: Elastic Scattering}
\label{app:collision}

In this section, we discuss the collision term for elastic scattering of two
identical particles $\chi$, i.e.\ the process
$\chi(p_1)\chi(p_2)\leftrightarrow\chi(p_3)\chi(p_4)$, with
$E_i=\sqrt{p_i^2+m_i^2}$. The full collision term is
\begin{equation}
    C(p_1,t) = -\frac{1}{2E_1}\int d\Pi_2\, d\Pi_3\, d\Pi_4\,
    (2\pi)^4\delta^4(p_1+p_2-p_3-p_4)\,|\mathcal{M}|^2
    \left[f_\chi(p_1,t)f_\chi(p_2,t)-f_\chi(p_3,t)f_\chi(p_4,t)\right].
    \label{app_eq:Cfull}
\end{equation}
It can always be split into a \textit{forward}
($\chi(p_1)\chi(p_2)\to\chi(p_3)\chi(p_4)$) and a \textit{backward}
($\chi(p_1)\chi(p_2)\leftarrow\chi(p_3)\chi(p_4)$) term,
$C=C_{\rm FW}+C_{\rm BW}$ with \cite{Du:2021jcj,Ala-Mattinen:2019mpa,Bringmann:2006mu}
\begin{align}
    C_{\rm FW}(p_1,t) &= -\frac{1}{2E_1}\int d\Pi_2\, d\Pi_3\, d\Pi_4\,
    (2\pi)^4\delta^4(p_1+p_2-p_3-p_4)\,|\mathcal{M}|^2
    f_\chi(p_1,t)f_\chi(p_2,t), \label{app_eq:Cfw}\\
    C_{\rm BW}(p_1,t) &= \frac{1}{2E_1}\int d\Pi_2\, d\Pi_3\, d\Pi_4\,
    (2\pi)^4\delta^4(p_1+p_2-p_3-p_4)\,|\mathcal{M}|^2
    f_\chi(p_3,t)f_\chi(p_4,t). \label{app_eq:Cbw}
\end{align}

This allows two ways of evaluating the collision term:
\begin{itemize}
    \item \textit{Method-1}: use the full collision term,
    Eq.~\ref{app_eq:Cfull}, where the difference of the distribution
    functions is taken inside the integral
    \cite{Hryczuk:2022gay,Beauchesne:2024zsq}.
    \item \textit{Method-2}: compute $C_{\rm FW}$ and $C_{\rm BW}$,
    Eqs.~\ref{app_eq:Cfw} and \ref{app_eq:Cbw}, separately and subtract
    them afterwards \cite{Ala-Mattinen:2022nuj,Yoon:2026rce}.
\end{itemize}
In both cases the integral over $p_2,p_3,p_4$ is nine-dimensional before the
delta function is used, which makes a direct numerical evaluation very
expensive. It can be reduced by a suitable parametrization of the momenta,
which we describe next for the forward and backward terms. The choice between
the two methods is discussed in Sec.~\ref{app:challenges}.

\subsubsection{Parametrization of the momenta}
\label{app:param}

\paragraph{Forward term.}
Following \cite{Hannestad:1995rs,Ala-Mattinen:2022nuj}, we choose
\begin{equation}
    \vec p_1=|\vec p_1|\,\hat z, \qquad
    \vec p_3=|\vec p_3|(0,\sin\theta,\cos\theta), \qquad
    \vec p_2=|\vec p_2|(\sin\alpha\sin\beta,\ \sin\alpha\cos\beta,\ \cos\alpha).
\end{equation}
Using the delta function, which removes four integrations, and the azimuthal
symmetry about $\vec p_1$, which removes one more, $C_{\rm FW}$ becomes
\begin{equation}
    C_{\rm FW}(p_1,t) = -\frac{1}{128\pi^4E_1}\int\frac{p_2^2\,dp_2}{E_2}
    \int\frac{p_3^2\,dp_3}{E_3}\,
    f_\chi(p_1,t)f_\chi(p_2,t)\,F(p_1,p_2,p_3)\,\Theta(p_1+p_2-p_3),
\end{equation}
where the angular integral is
\begin{equation}
    F(p_1,p_2,p_3)=2\int_{-1}^{1}d\cos\theta\int_{-1}^{1}d\cos\alpha
    \left[\frac{|\mathcal{M}|^2}{\sqrt{a\cos^2\alpha+b\cos\alpha+c}}\,
    \Theta\!\left(a\cos^2\alpha+b\cos\alpha+c\right)\right].
    \label{app_eq:F_fw}
\end{equation}
The coefficients $a$, $b$ and $c$ are
\begin{align}
    a &= 4p_2^2\left(-p_1^2-p_3^2+2p_1p_3\cos\theta\right),\\
    b &= 4p_2\left(p_1-p_3\cos\theta\right)\,\Delta_{\rm FW},\\
    c &= 4p_2^2p_3^2\left(1-\cos^2\theta\right)-\Delta_{\rm FW}^2,
\end{align}
with
\begin{equation}
    \Delta_{\rm FW}\equiv m_1^2+m_2^2+m_3^2-m_4^2
    +2\left(E_1E_2-E_1E_3-E_2E_3\right)+2p_1p_3\cos\theta .
\end{equation}

\paragraph{Backward term.}
Similarly, for $C_{\rm BW}$ we use \cite{Hannestad:1995rs,Ala-Mattinen:2022nuj}
\begin{equation}
    \vec p_1=|\vec p_1|\,\hat z, \qquad
    \vec p_3=|\vec p_3|(0,\sin\theta,\cos\theta), \qquad
    \vec p_4=|\vec p_4|(\sin\alpha\sin\beta,\ \sin\alpha\cos\beta,\ \cos\alpha),
\end{equation}
which gives
\begin{equation}
    C_{\rm BW}(p_1,t) = \frac{1}{128\pi^4E_1}\int\frac{p_3^2\,dp_3}{E_3}
    \int\frac{p_4^2\,dp_4}{E_4}\,
    f_\chi(p_3,t)f_\chi(p_4,t)\,\tilde F(p_1,p_3,p_4)\,\Theta(p_3+p_4-p_1),
\end{equation}
with
\begin{equation}
    \tilde F(p_1,p_3,p_4)=2\int_{-1}^{1}d\cos\theta\int_{-1}^{1}d\cos\alpha
    \left[\frac{|\mathcal{M}|^2}{\sqrt{\tilde a\cos^2\alpha+\tilde b\cos\alpha+\tilde c}}\,
    \Theta\!\left(\tilde a\cos^2\alpha+\tilde b\cos\alpha+\tilde c\right)\right].
    \label{app_eq:F_bw}
\end{equation}
The coefficients are
\begin{align}
    \tilde a &= 4p_4^2\left(-p_1^2-p_3^2+2p_1p_3\cos\theta\right),\\
    \tilde b &= -4p_4\left(p_1-p_3\cos\theta\right)\,\Delta_{\rm BW},\\
    \tilde c &= 4p_3^2p_4^2\left(1-\cos^2\theta\right)-\Delta_{\rm BW}^2,
\end{align}
with
\begin{equation}
    \Delta_{\rm BW}\equiv m_1^2-m_2^2+m_3^2+m_4^2
    +2\left(E_3E_4-E_1E_3-E_1E_4\right)+2p_1p_3\cos\theta .
\end{equation}

\subsubsection{Angular integrals}
\label{app:angular}

The angular integrals $F$ and $\tilde F$ have the same structure and differ only
in their coefficients, so it suffices to treat $F(p_1,p_2,p_3)$. For a constant
matrix element $|\mathcal{M}|^2$, the integral over $\cos\alpha$ can be done
analytically \cite{Hannestad:1995rs,Ala-Mattinen:2022nuj}, giving
\begin{equation}
    F = |\mathcal{M}|^2\int d\cos\theta\,\frac{2\pi}{\sqrt{-a}}\,
    \Theta\!\left(b^2-4ac\right).
\end{equation}
The remaining $\cos\theta$ integral is also analytic. Its limits are set by the
roots of $b^2-4ac=0$, denoted $\cos\theta_{\rm min}$ and $\cos\theta_{\rm max}$:
the lower limit is $\sup[-1,\cos\theta_{\rm min}]$ and the upper limit is
$\inf[1,\cos\theta_{\rm max}]$ \cite{Hannestad:1995rs}. The resulting angular
integrals are then inserted into the collision terms. For a momentum-dependent
$|\mathcal{M}|^2$, the angular integrals are instead evaluated numerically.

\subsection{Energy Conservation and the Choice of Method}
\label{app:challenges}

A crucial issue in the numerical evaluation of the collision term is energy
conservation. In the elastic process
$\chi(p_1)\chi(p_2)\leftrightarrow\chi(p_3)\chi(p_4)$,
\begin{equation}
    E_1+E_2=E_3+E_4 .
    \label{app_eq:Econs}
\end{equation}

\paragraph{Problem with splitting the collision term (Method-2).}
In the forward term, $p_1$, $p_2$ and $p_3$ (hence $E_1$, $E_2$ and $E_3$) are
specified, and Eq.~\ref{app_eq:Econs} fixes $E_4$. The forward term only needs
$f_\chi(E_1)$ and $f_\chi(E_2)$, which are available exactly on the momentum
grid. The backward term, however, needs $f_\chi(E_3)$ and $f_\chi(E_4)$. If both
terms are to be evaluated on the same $(p_2,p_3)$ grid, $f_\chi(E_3)$ is read
from the grid, but $f_\chi(E_4)$ must be evaluated at $E_4=E_1+E_2-E_3$, which
generally does not coincide with a grid point. Conversely, in the $(p_3,p_4)$
parametrization it is $E_2$ that is fixed by energy conservation.

Making $E_4$ (or $E_2$) land exactly on a grid point would require an extremely
fine momentum grid, which is computationally expensive. The integrals over
$(p_2,p_3)$ in the forward case, or $(p_3,p_4)$ in the backward case, are usually
performed as nested loops that call $F$ and $\tilde F$ at every iteration, and
for such a fine grid these loops take an enormous amount of time. To avoid this,
one can precompute $F$ and $\tilde F$ on the grid and simply look them up inside
the loops. However, this introduces a memory problem: each of $F$ and $\tilde F$
depends on three momenta and on $x$, so with $100$ points in $x$ and $1000$
points per momentum, a single double-precision table requires
\begin{equation}
    100\times(1000)^3\times 8~\text{bytes}=8\times10^{11}~\text{bytes}\approx 800~\text{GB},
\end{equation}
which is practically prohibitive. One is therefore forced to use a coarser grid
than the one required, so $E_4$ (or $E_2$) generally does not land on a grid
point, and energy conservation is violated.

\paragraph{Method-1.}
To avoid these issues, we do not split the collision term into forward and
backward parts. We follow Method-1, where the difference of the distribution
functions is taken before the final integration. The delta function can then be
handled with either the forward or the backward parametrization. Using the
forward one, the collision term becomes
\begin{equation}
    \begin{aligned}
        C_{\rm full}(p_1,t) = -\frac{1}{128\pi^4E_1}
        \int\frac{p_2^2\,dp_2}{E_2}\int\frac{p_3^2\,dp_3}{E_3}\,
        F(p_1,p_2,p_3)\,\Theta(p_1+p_2-p_3)
        \Big[ & f_\chi(p_1,t)f_\chi(p_2,t)\\
              & - f_\chi(p_3,t)\,f_\chi(p_4,t)\Big],
    \end{aligned}
    \label{app_eq:Cfull_fw}
\end{equation}
where $p_4$ is fixed by $E_4=E_1+E_2-E_3$. This value is required by the
energy-conserving delta function. To evaluate $f_\chi(p_4,t)$, we interpolate the
distribution function in energy (or momentum) at each time step. The
interpolant gives $f_\chi$ at any point, so $E_4$ need not lie on the grid. A
fine grid is therefore not needed to catch $E_4$, and a reasonable number of
grid points suffices. Since only $F$ is needed (and not $\tilde F$), the memory
requirement is also halved.

\subsection{Transformation of Variables}
\label{app:variable_transformation}

Having specified the collision term in Sec.~\ref{app:collision}, we now
simplify the left-hand side of Eq.~\ref{app:f_eq_phy}; the number-density
equation, Eq.~\ref{app:n_eq_phy}, is treated in Sec.~\ref{app:final_form_BE}.
The Hubble-friction term only accounts for the redshift due to expansion and
carries no information about the interactions, so it is convenient to change
variables such that it is removed.

The distribution function depends on the time $t$ and the physical momentum $p$.
Since $p$ redshifts with time, solving the equation numerically in these
variables is challenging. To remove the redshift, we use the comoving momentum
$k = a p$, which is constant for free-streaming particles. We also replace the
time $t$ by the dimensionless variable $x = m/T$, where $T$ is the temperature of
the thermal bath and $m$ is an arbitrary reference mass scale. Denoting the distribution function
in the new variables by the same symbol, we write
\begin{equation}
    f(t,p) \equiv f(x,k).
\end{equation}
The Liouville operator is then transformed from $(t,p)$ to $(x,k)$ as
\begin{equation}
    \begin{aligned}
        \left.\frac{\partial}{\partial t}\right|_p - Hp \left.\frac{\partial}{\partial p}\right|_t
        & = \left.\frac{dx}{dt}\right|_p \frac{\partial}{\partial x}
          + \left.\frac{dk}{dt}\right|_p \frac{\partial}{\partial k}
          - Hp \left.\frac{dk}{dp}\right|_t \frac{\partial}{\partial k} \\
        & = \left.\frac{dx}{dt}\right|_p \frac{\partial}{\partial x}
          + \dot{a}p \frac{\partial}{\partial k}
          - Hpa \frac{\partial}{\partial k} \\
        & = \left.\frac{dx}{dt}\right|_p \frac{\partial}{\partial x},
    \end{aligned}
    \label{app:t_p_to_x_k}
\end{equation}
where the last step uses $\dot a = Ha$, so the two $\partial/\partial k$ terms
cancel exactly.

It remains to compute $dx/dt$, which requires the time evolution of the
temperature. Assuming entropy conservation, $T\propto g_{*,s}^{-1/3}a^{-1}$,
and assuming the Universe is radiation dominated, the Hubble rate is
\cite{Ala-Mattinen:2022nuj}
\begin{equation}
    T \propto g_{*,s}^{-1/3}\, a^{-1}, \qquad
    H = \frac{\dot a}{a} = \sqrt{\frac{\rho_R}{3M_P^2}}
      = \sqrt{\frac{\pi^2 g_{*,\rm rel}}{90\,M_P^2}}\;T^2,
    \label{app:rad_dom_assum}
\end{equation}
where $M_P$ is the reduced Planck mass, and $g_{*,s}$ and $g_{*,\rm rel}$ are
the entropic and relativistic effective degrees of freedom, respectively.
Taking the logarithmic time derivative of the first relation gives
\begin{equation}
    -\frac{1}{T}\frac{dT}{dt}
    = \frac{H}{1 + \frac{T}{3g_{*,s}}\frac{dg_{*,s}}{dT}}
    = \frac{H(x)}{1 - \frac{x}{3g_{*,s}}\frac{dg_{*,s}}{dx}}
    \equiv \tilde H(x),
    \label{app:T_evolution}
\end{equation}
and therefore
\begin{equation}
    \left.\frac{dx}{dt}\right|_p = -x\,\frac{1}{T}\frac{dT}{dt} = x\tilde H(x).
    \label{app:t_to_x}
\end{equation}

To make the momentum variable dimensionless, we rescale the comoving momentum as
\begin{equation}
    \xi = \frac{k}{a_0T_0} = \frac{a p}{a_0T_0}
    = \left(\frac{g_{*,s}(T_0)}{g_{*,s}(T)}\right)^{1/3}\frac{p}{T},
    \label{app:p_to_xi}
\end{equation}
where $a_0$ and $T_0$ are the scale factor and temperature at some reference
point. Since $\xi$ differs from $k$ only by a constant, holding $k$ fixed is
equivalent to holding $\xi$ fixed, and Eq.~\ref{app:t_p_to_x_k} holds with
$k\to\xi$. The inverse relation,
\begin{equation}
    p_i = \xi_i\,T\left(\frac{g_{*,s}(T)}{g_{*,s}(T_0)}\right)^{1/3},
    \qquad E_i = \sqrt{p_i^2+m_{\chi, i}^2},
    \label{app:xi_to_p}
\end{equation}
is what we need to express the collision term in the new variables.

With these variables, the Liouville operator of Eq.~\ref{app:f_eq_phy}
reduces, through Eqs.~\ref{app:t_p_to_x_k} and \ref{app:t_to_x}, to a
single derivative in $x$ at fixed $\xi$. The Boltzmann equation then reads
\begin{equation}
    x\tilde H(x)\,\frac{d}{dx}f_\chi(x,\xi_1)=\mathcal{C}[f_\chi],
    \label{app:f_eq_com}
\end{equation}
where $\mathcal{C}[f_\chi]$ is the full right-hand side of
Eq.~\ref{app:f_eq_phy}, with all momenta and energies expressed through
$(x,\xi_i)$ using Eq.~\ref{app:xi_to_p}. The partial derivative has become an
ordinary one because $\xi$ is held fixed.

\subsection{Final Form of the Boltzmann Equation}
\label{app:final_form_BE}

\subsubsection{Distribution-function level}
Combining the transformed Liouville operator, Eq.~\ref{app:f_eq_com}, with
the collision term of Method-1 in the forward parametrization,
Eq.~\ref{app_eq:Cfull_fw}, and expressing all momenta through $\xi_i$ using
Eq.~\ref{app:xi_to_p}, the Boltzmann equation in terms of the scaled comoving
momentum becomes
\begin{equation}
    \begin{aligned}
        \frac{d f_\chi(x,\xi_1)}{dx}
        = -\frac{1}{x\tilde H(x)}\,\frac{1}{128\pi^4E_1}
        \left(\frac{m}{x}\right)^{6} g_{\rm fact}(x)
        \int\frac{\xi_2^2\,d\xi_2}{E_2}\int\frac{\xi_3^2\,d\xi_3}{E_3}\,
        F(x,\xi_1,\xi_2,\xi_3)\, \\
        \Big[ f_\chi(x,\xi_1)\,f_\chi(x,\xi_2)
               - f_\chi(x,\xi_3)\,f_\chi(x,\xi_4)\Big],
    \end{aligned}
    \label{app_eq:f_eq_final}
\end{equation}
where $m$ is the reference mass scale of $x=m/T$, and
$g_{\rm fact}(x)=\left[g_{*,s}(x)/g_{*,s}(x_0)\right]^{2}$ arises from the
Jacobian $p_i=\xi_i T\,[g_{*,s}(x)/g_{*,s}(x_0)]^{1/3}$. The energies are
$E_i=\sqrt{p_i^2+m_{\chi, i}^2}$, and $\xi_4$ is fixed by energy conservation,
$E_4=E_1+E_2-E_3$. Since this value generally does not lie on the $\xi$ grid,
$f_\chi(x,\xi_4)$ is obtained by interpolation, as discussed in
Sec.~\ref{app:challenges}.

In writing Eq.~\ref{app_eq:f_eq_final} we neglected the derivative of
$g_{*,s}$ with respect to $x$, so that $\tilde H(x)=H(x)$. This is a good
approximation because $g_{*,s}$ is piecewise constant and varies appreciably
only near thresholds such as the QCD transition. The values of $g_{*,s}$
themselves are retained, through $g_{\rm fact}$ and $H(x)$.

\subsubsection{Number-density equation}
The collision term of Eq.~\ref{app:n_eq_phy}, for elastic scattering of
identical particles $\chi(p_1)\chi(p_2)\leftrightarrow\chi(p_3)\chi(p_4)$, is
\begin{equation}
    \mathcal{C}[n_\chi] = \int d\Pi_1\, d\Pi_2\, d\Pi_3\, d\Pi_4\,
    (2\pi)^4\delta^4(p_1+p_2-p_3-p_4)\,|\mathcal{M}|^2
    \left[f_\chi(p_1)f_\chi(p_2)-f_\chi(p_3)f_\chi(p_4)\right],
    \label{app_eq:Cn}
\end{equation}
so that Eq.~\ref{app:n_eq_phy} reads
$\dot n_\chi+3Hn_\chi=-\mathcal{C}[n_\chi]$. We now show that
$\mathcal{C}[n_\chi]=0$. Write $\mathcal{C}[n_\chi]=I_1-I_2$, where $I_1$ and
$I_2$ are the integrals of the $f_\chi(p_1)f_\chi(p_2)$ and
$f_\chi(p_3)f_\chi(p_4)$ terms, respectively. In $I_2$ all four momenta are
integrated over, so they are dummy variables and we may relabel
$(p_1,p_2)\leftrightarrow(p_3,p_4)$. This is harmless because:
\begin{itemize}
    \item the measure $d\Pi_1 d\Pi_2 d\Pi_3 d\Pi_4$ is symmetric under the
    exchange, since all four particles have the same mass;
    \item the delta function is symmetric,
    $\delta^4(p_1+p_2-p_3-p_4)=\delta^4(p_3+p_4-p_1-p_2)$;
    \item the squared amplitude is symmetric under the exchange of initial and
    final states, $|\mathcal{M}|^2_{12\to34}=|\mathcal{M}|^2_{34\to12}$, which
    follows from time-reversal invariance. For elastic scattering of identical
    particles this is satisfied.
\end{itemize}
After the relabeling, $I_2$ becomes identical to $I_1$, hence
\begin{equation}
    \mathcal{C}[n_\chi]=I_1-I_2=0 .
    \label{coll_n_sc}
\end{equation}
Physically, each elastic collision removes two $\chi$ particles and creates two,
so the total number is unchanged. Eq.~\ref{app:n_eq_phy} then reduces to
\begin{equation}
    \frac{dn_\chi}{dt}+3Hn_\chi=0
    \quad\Rightarrow\quad n_\chi\propto\frac{1}{a^3},
    \label{app_eq:n_sol}
\end{equation}
i.e.\ the number density only redshifts.

\subsubsection{Energy-density equation}
For the energy density, the collision term carries the additional weight $E_1$,
\begin{equation}
    \mathcal{C}[\rho_\chi] = \int d\Pi_1 d\Pi_2 d\Pi_3 d\Pi_4\,
    (2\pi)^4\delta^4(p_1+p_2-p_3-p_4)\,|\mathcal{M}|^2\,E_1
    \left[f_\chi(p_1)f_\chi(p_2)-f_\chi(p_3)f_\chi(p_4)\right].
\end{equation}
Since $E_1$ is not invariant under relabeling, we symmetrize each term
separately, using the symmetry of the measure, the delta function and
$|\mathcal{M}|^2$. In the first term, exchanging $p_1\leftrightarrow p_2$ gives
$E_1\to(E_1+E_2)/2$. In the second term, exchanging
$(p_1,p_2)\leftrightarrow(p_3,p_4)$ and then $p_3\leftrightarrow p_4$ turns
$E_1f_\chi(p_3)f_\chi(p_4)$ into $\tfrac12(E_3+E_4)f_\chi(p_1)f_\chi(p_2)$.
Therefore
\begin{equation}
    \mathcal{C}[\rho_\chi]=\int d\Pi_1 d\Pi_2 d\Pi_3 d\Pi_4\,
    (2\pi)^4\delta^4(p_1+p_2-p_3-p_4)\,|\mathcal{M}|^2\,
    \frac{E_1+E_2-E_3-E_4}{2}\,f_\chi(p_1)f_\chi(p_2)=0,
\end{equation}
because the delta function enforces $E_1+E_2=E_3+E_4$. Physically, an elastic
collision leaves the total energy unchanged. In the relativistic regime
($P_\chi=\rho_\chi/3$) this gives
\begin{equation}
    \frac{d\rho_\chi}{dt}+4H\rho_\chi=0
    \quad\Rightarrow\quad \rho_\chi\propto\frac{1}{a^4},
    \label{app_eq:rho_sol_rel}
\end{equation}
whereas in the non-relativistic regime ($P_\chi\ll\rho_\chi$)
\begin{equation}
    \frac{d\rho_\chi}{dt}+3H\rho_\chi=0
    \quad\Rightarrow\quad \rho_\chi\propto\frac{1}{a^3}.
    \label{app_eq:rho_sol_nr}
\end{equation}
Thus the energy density is only diluted by the expansion.

\subsubsection{A Final Check}
\begin{figure}[H]
    \centering
    \includegraphics[width=0.6\linewidth]{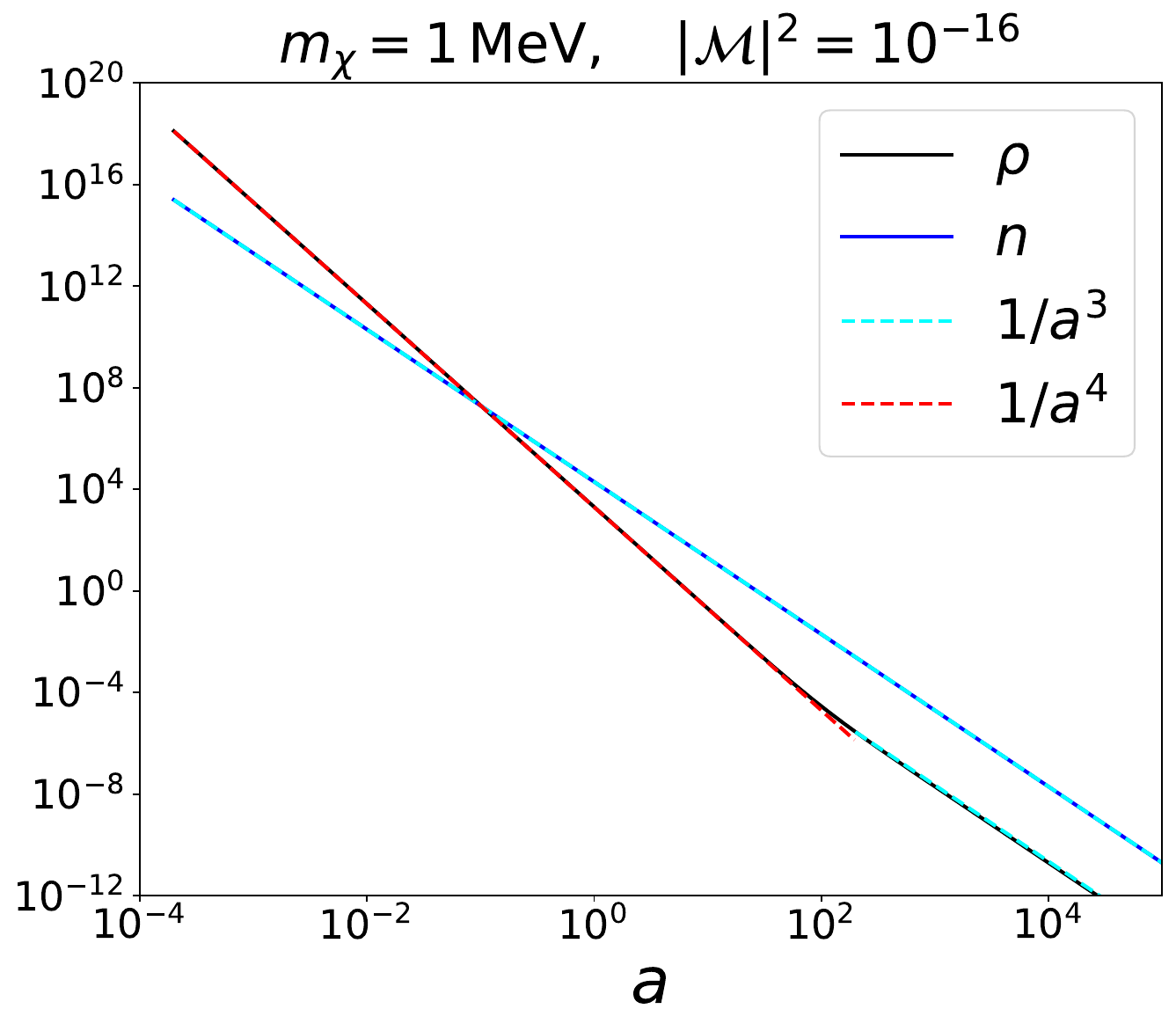}
    \caption{Evolution of the energy density (black) and number density (gray),
    computed from the solution of Eq.~\ref{app_eq:f_eq_final}. The dashed
    curves show the expected scalings: $1/a^3$ (cyan), the solution of
    Eqs.~\ref{app_eq:n_sol} and \ref{app_eq:rho_sol_nr}, and $1/a^4$ (red),
    the solution of Eq.~\ref{app_eq:rho_sol_rel}.}
    \label{fig:f_ncheck}
\end{figure}

We conclude by checking the consistency between the solution of the
distribution-function-level equation and the number (energy) density equations.
After solving Eq.~\ref{app_eq:f_eq_final}, we compute the number and energy
densities as
$n_\chi=\int\frac{d^3p}{(2\pi)^3}\,f_\chi$ and
$\rho_\chi=\int\frac{d^3p}{(2\pi)^3}\,E\,f_\chi$.
Fig.~\ref{fig:f_ncheck} shows $n_\chi$ and $\rho_\chi$ obtained this way, for
$m_\chi=1~{\rm MeV}$ and $|\mathcal{M}|^2=10^{-16}$. The dashed curves show the
expected scalings from Eqs.~\ref{app_eq:n_sol}, \ref{app_eq:rho_sol_rel}
and \ref{app_eq:rho_sol_nr}. The numerical results agree excellently with
these scalings, as expected, which confirms the correctness of our numerics.
It is worth mentioning that, our code has also been cross-checked against the recently published code BEST~\cite{Yoon:2026rce}. In that work, quantum effects were considered with $H=0$. Our results agree exactly with theirs in the same limit.

\section{Decoupling time for calculating free-streaming length}
\label{app:B}

\begin{figure}[h]
\includegraphics[scale=0.225]{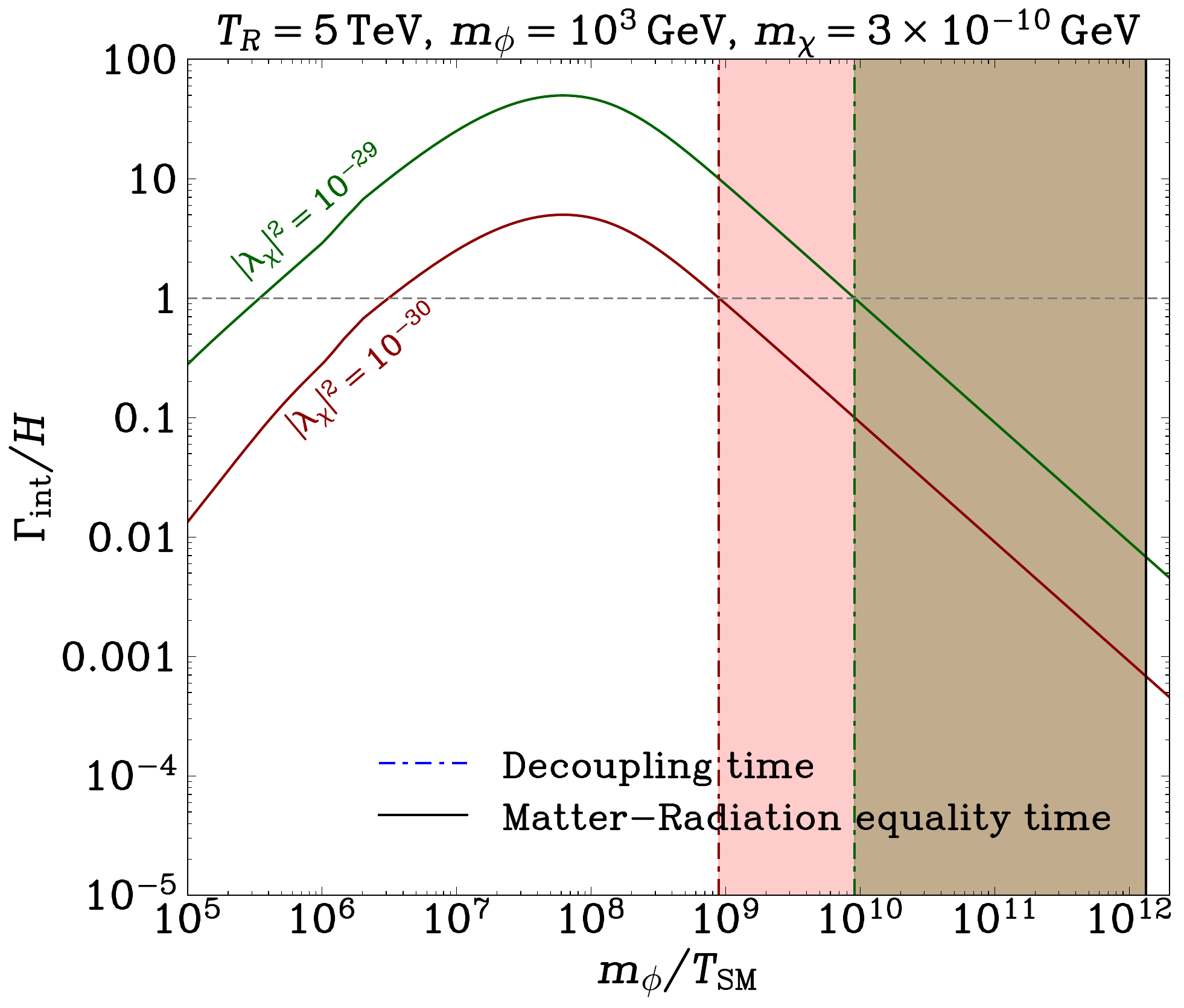}
\includegraphics[scale=0.225]{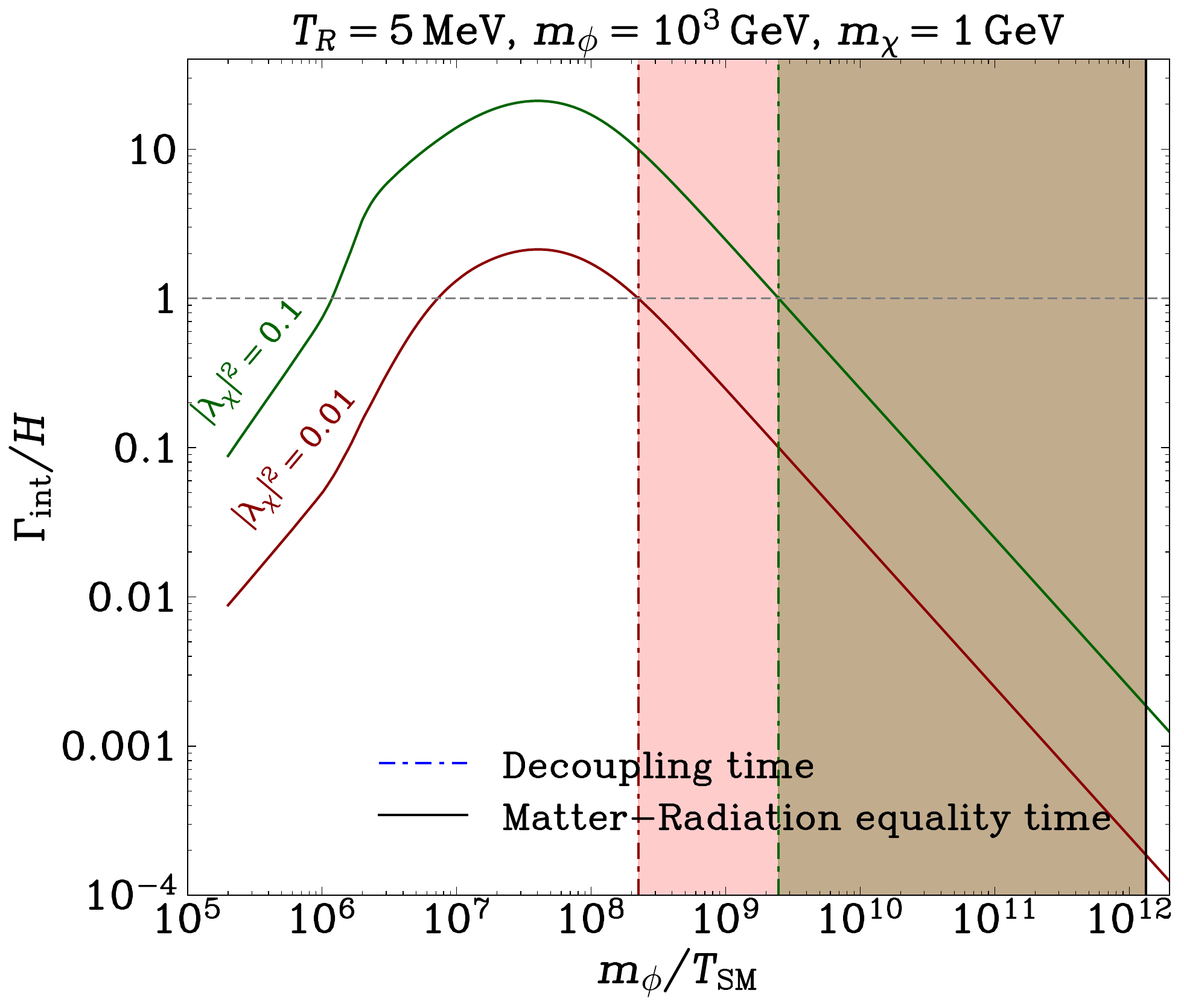}
\caption{Ratio of the self-scattering rate to the Hubble rate for reheating
temperatures $T_R = 5\times10^6~{\rm GeV}$ (left panel) and
$T_R = 5~{\rm MeV}$ (right panel). The horizontal dashed line marks
$\Gamma_{\rm int}=H$, the vertical dashed lines mark the decoupling times,
and the black line marks matter--radiation equality. The yellow shaded region shows the
free-streaming duration of DM particles .}
\label{fig:GammaByH}
\end{figure}

The co-moving free-streaming length of the DM particles is computed from the time
when elastic scattering within the secluded sector stops until the matter--radiation
equality epoch. So we need to know the decoupling time of elastic scatterings. We obtain it by comparing the
self-interaction rate, $\Gamma_{\rm int}\sim n_\chi\lambda_\chi^2/E_\chi^2$ as in
Eq.~\ref{eq:Gamma} with the Hubble rate, i.e. $H\sim T_{\rm SM}^2\propto a^{-2}$.
Fig.~\ref{fig:GammaByH} shows $\Gamma_{\rm int}/H$ as a function of
$m_\phi/T_{\rm SM} (\propto a)$ for the two benchmark reheat temperatures.

When DM particles are relativistic, $n_\chi\propto a^{-3}$ and $E_\chi\propto a^{-1}$, $\Gamma_{\rm int}\propto a^{-1}$ and hence
\begin{equation}
    \frac{\Gamma_{\rm int}}{H}\propto a ,
\end{equation}
which implies scatterings becoming efficient over time. For \ $\Gamma_{\rm int}/H \gtrsim 1$, scatterings become fast compared to the expansion and DM particles thermalize. The first transition of $\Gamma_{\rm int}/H$ to above unity justifies treating DM particles as a thermalized
species at later times.

Once DM particles become non-relativistic, i.e. $E_\chi\sim m_\chi$, the elastic cross-section saturates at
$\sigma_{\chi \chi}\sim\lambda_\chi^2/m_\chi^2$ for given values of $\lambda_\chi$ and $m_\chi$. The interaction rate then redshifts as the DM number density, i.e. $\Gamma_{\rm int}\propto n_\chi\propto a^{-3}$, which sets
\begin{equation}
    \frac{\Gamma_{\rm int}}{H}\propto a^{-1}.
\end{equation}
The ratio peaks near this transition and then falls, crossing unity for the second time from above. For the left panel of Fig.~\ref{fig:GammaByH}, this occurs at $m_\phi/T_{\rm SM}\simeq 9\times10^{8}$ for $\lambda_\chi^2=10^{-30}$ and at $m_\phi/T_{\rm SM}\simeq 9\times10^{9}$ for $\lambda_\chi^2=10^{-29}$. For the right panel, the crossings occur at $m_\phi/T_{\rm SM}\simeq 2\times10^{8}$ for $\lambda_\chi^2=0.01$ and at $m_\phi/T_{\rm SM}\simeq 2\times10^{9}$ for $\lambda_\chi^2=0.1$. These second
crossings mark the decoupling times (red and green dashed lines): beyond them, DM particles scatter less than once per Hubble time and can no longer exchange energy or momentum efficiently within the dark sector. The DM therefore free-streams from decoupling until matter--radiation equality, and the shaded regions in Fig.~\ref{fig:GammaByH} show the duration of this free-streaming
epoch for the chosen parameters. The lower limit of the free-streaming integral is thus set by the condition $\Gamma_{\rm int}(a_{\rm dec})=H(a_{\rm dec})$.


\end{document}